\documentclass[11pt]{book}
\usepackage{Wiley-AuthoringTemplate}
\usepackage{chapterbib}
\usepackage[sectionbib,authoryear]{natbib}

\makeindex

\begin{document}

\frontmatter

\mainmatter

\chapter[Transport in helical fluid turbulence]{Transport in helical fluid turbulence\protect\footnote{Visiting researcher at Nordic Institute for Theoretical Physics (NORDITA).}}

\author*[1]{Nobumitsu Yokoi}

\address[1]{\orgdiv{Institute of Industrial Science}, 
\orgname{University of Tokyo}, 
\postcode{153-8505}, \countrypart{Tokyo}, 
     \city{Meguro}, \street{Komaba}, \country{Japan}}%



\address*{Corresponding Author: Nobumitsu Yokoi; \email{nobyokoi@iis.u-tokyo.ac.jp}}

\def\NY#1{{\textcolor{magenta}{#1}}} 

\maketitle



\begin{abstract}{}
Kinetic helicity (hereafter helicity) is defined by the correlation between the velocity and the flow-aligned vorticity. Helicity, as well as energy, is an inviscid invariant of the hydrodynamic equations. In contrast to energy, a measure of the turbulent intensity, turbulent helicity, representing right- and left-handed twist associated with a fluctuating motion, provides a measure of the structural or topological property of the fluctuation. The helicity effect on the turbulent transport can be analytically obtained in the framework of the multiple-scale renormalized perturbation expansion theory through the inclusion of the non-reflectionally-symmetric part for the lowest-order (homogeneous and isotropic) velocity correlation. The physical significance of the helicity-related contribution to the momentum transport is explained. By utilizing the analytical expression of the Reynolds stress, a turbulence model with helicity effect incorporated (helicity model) is constructed. This helicity model is applied to a swirling flow to show its validity in describing the prominent properties of the flow. In addition to the transport suppression, inhomogeneous helicity coupled with a rotation can induce a large-scale flow. The results of direct numerical simulations (DNSs) confirming the global flow generation by helicity will be also reviewed, followed by several possible applications in geo- and astro-physical flow phenomena.
\end{abstract}


\section{Introduction}
Kinetic helicity is a measure of broken mirror-symmetry and represents the geometrical and topological properties of fluid turbulence. The presence of helicity alters the dynamical and statistical properties of turbulence. Beyond fluid dynamics, helicity plays a role in diverse areas of research including DNA and bio-chemistry, ventral nodal flow in generation of left-right asymmetry, vortex-entanglement in quantum fluids, chirality in neutrino left-right skewness asymmetry in the universe, zonal flow generation in fusion plasma devices etc.

	The first discussions on the importance of helicity were likely to be evoked in the context of magnetic-field generation: dynamo. If the fluid motion is helical, magnetic field frozen into the fluid can be also twisted. Then, the mean electric-current density configuration parallel or antiparallel to the original mean magnetic flux tube can be generated. Assuming a cyclonic fluid motion, which is related to the cyclone and anticyclone observed in the atmosphere, in small scales, \citet{par1955}  proposed a mechanism in which a azimuthal or toroidal magnetic field is deformed to become a dipole or poloidal magnetic field and vice versa. This is the $\alpha$ dynamo which is considered to be one of the most relevant processes in the stellar migratory dynamo cycle. In this dynamo, helicity in turbulence is an essential ingredient. An inhomogeneous velocity fluctuation ${\bf{u}}'$ along the large-scale magnetic field ${\bf{B}}$ induces an magnetic-field fluctuation ${\bf{b}}'$. In the presence of kinetic helicity in turbulence, $\langle {{\bf{u}}' \cdot (\nabla \times {\bf{u}}')} \rangle$, an electromotive force component, $\langle {{\bf{u}}' \times {\bf{b}}'} \rangle$, parallel or anti-parallel to the large-scale magnetic field can be induced depending on the sign of turbulent helicity ($\langle {\cdots} \rangle$: ensemble average). This gives marked contrast with the turbulent energy, which always induces an electromotive force anti-parallel to the large-scale electric-current density, leading to the enhancement of magnetic diffusivity due to turbulence. In the presence of the current helicity in turbulence, $\langle {{\bf{b}}' \cdot (\nabla \times {\bf{b}}')} \rangle$, mediated by a part of the fluctuating Lorentz force $(\nabla \times {\bf{b}}') \times {\bf{B}}$, an electromotive force can be induced in the direction parallel or antiparallel to the large-scale magnetic field ${\bf{B}}$. For the kinetic-helicity effect in dynamo as well as the other helicities effects, the reader is referred to textbooks and review papers on dynamos \citep{mof1978,par1979,kra1980,bra2005,yok2013,mof2019,tob2021}.

	Since helicity is a measure of broken mirror-symmetry, a prerequisite for non-zero helicity is an element that breaks mirror-symmetry. Rotation is one of such elements that causes breakage of mirror-symmetry. As we see later in the transport equation of turbulent helicity (\S~4.2), rotation coupled with inhomogeneity provides turbulence with a finite helicity through the flux across the boundary. In geophysical and astrophysical flows, rotation and density stratification are basic ingredients of the system. A strong stratification (low Froude number $Fr = U/(N_{\rm{BV}}h) \lesssim 1$ with $U$ being the horizontal velocity scale, $h$ the vertical length scale, and $N_{\rm{BV}}$ the Brunt--V\"{a}is\"{a}l\"{a} frequency) causes the fluid to be deflected almost purely horizontally, and flow becomes two-dimensional. Even with this two-dimensionality in the large-scale flow structure, small-scale turbulent motions are usually more complex and three-dimensional. For example, generation of helicity due to rotation and stratification is a concept dating back to \citet{ste1966} and to \citet{hid1989} who demonstrated how it could occur. In the presence of inhomogeneity along the rotation axis, turbulence in a stratified fluid can be helical \citep{mar2013a,mar2013b}. 
	
	Related to the dimensionality of turbulence, another interesting point is the helicity effect in two-dimensional with three-component configuration, 2D3C. In the presence of a strong rotation, depending on the boundary conditions, the flow becomes quasi 2D3C state, where the helicity invariants are still present. From the viewpoint of the helicity effect, which we present in \S~\ref{sec:3.3}, helicity inhomogeneity in space is essential. Inhomogeneity of helicity cannot be present in the third direction perpendicular to the 2D or horizontal plane in the 2D3C configuration. On the other hand, helicity inhomogeneity in the horizontal plane is possible. This horizontally inhomogeneous helicity coupled with the perpendicular mean absolute vorticity (associated with the rotation and non-uniform horizontal flows) may contribute to the horizontal momentum transfer through the Reynolds stress. These features of the 2D3C configuration would be interesting subject to explore in the future as well as the cascade direction and scalings of energy and helicity associated with the constraints involving invariants other than energy.

	In the fundamental theoretical studies of turbulence, how and how much  helicity influences energy cascade and dissipation have been often investigated using homogeneous isotropic turbulence (HIT). By constructing statistical mechanics for the truncated system of turbulence equation, \citet{kra1973} postulated a cascade of helicity towards the small scales. Since the classical work by \citet{and1977}, the influence of helicity on the evolution of turbulent spectra has been examined. In addition, the relationship between the high helicity region and low dissipation region in turbulence has been studied by \citet{rog1987} in several turbulent flow geometries. Recent progress in our understanding of helical process in fluid turbulence has been made in various ways. Introducing the helical decomposition of the velocity, \citet{wal1993} argued energy and helicity cascades with two classes of triad interaction of turbulence depending on whether the longest legs of triad are helical modes of the same or opposite sign. Using direct numerical simulation (DNS) of HIT 3D turbulence, it was shown in \citet{che2003} that the strong fluxes of positive and negative helical modes cancel with each other at high wave numbers, leading to a return to full isotropy at small scale. By examining the sub-categories of non-linear interaction, it has been found that the different subsets of interaction conserve the different combinations of the turbulent energy and helicity evolutions, leading to the possibility of partial inverse cascade of the energy of helical turbulence \citep{bif2012,ale2018}. By using DNSs of HIT, Taylor--Green (TG) flow with laboratory experiments of mirror-symmetric HIT and of chiral von K\'{a}rm\'{a}n flows, the relationship between the linking number of tracer trajectories and helicity has been established. This paves the way for experimentally measuring helicity on a firm mathematical basis \citep{ang2021}. An outline of the developments in homogeneous turbulence studies are briefly reviewed in \citet{pou2022}.

	In this chapter, we will not deal with the problem of cascade and dissipation in helical homogeneous turbulence, rather we focus our attention on the transport in helical inhomogeneous turbulence. What is the role of turbulent kinetic helicity in the dynamics of large-scale or mean velocity and vorticity? We present how to theoretically tackle this problem in inhomogeneous turbulence.

	From the viewpoint of transport, the primary effect of turbulence is to drastically enhance the effective transport. The eddy viscosity, eddy diffusivity, turbulent resistivity, etc. are the representative transport enhancement effects, which are determined by the intensity and timescale of turbulence. At the same time, in some cases where symmetry is broken in turbulence, the possibility of suppressing  the effective transport arises also due to turbulence. In this suppression process, quantity related to the breakage of symmetry plays a key role in counterbalancing the transport enhancement. As will be shown in the later sections, helicity is expected to play a crucial role in transport suppression in non-mirror-symmetric fluid turbulence. 

	In contrast to homogeneous turbulence, where energy injection by external forcing is indispensable for sustaining turbulence, in inhomogeneous turbulence, turbulent energy and helicity are naturally provided by the system itself through the production mechanisms of these quantities arising from the large-scale inhomogeneous fields, such as mean velocity shear, vorticity, temperature gradient, etc., and the fluxes from the boundaries. Mean field are determined by turbulence through the turbulent transport coefficients, and turbulence fields are determined by the production and transport rates directly linked to the large-scale inhomogeneities. In order to treat this nonlinear mean--turbulence interaction in a consistent manner, we have to simultaneously consider and solve the mean and turbulence fields. Direct numerical simulations (DNSs) of real inhomogeneous turbulence at realistic parameters for the geophysical and astrophysical phenomena with complex geometry and boundaries are just impossible. In this situation, turbulence modeling approach provides a very powerful tool for analyzing the real turbulent flows. We deal with how to incorporate such suppression effects of helicity into turbulence modeling.

	The organization of this chapter is as follows. First in \S~\ref{sec:2}, we present the definitions and basic properties of kinetic helicity. In \S~\ref{sec:3}, how to theoretically treat helicity effect in inhomogeneous turbulence at very high Reynolds number will be shown. One of the key procedures in the theoretical formulation for helical turbulence is to adopt the non-mirror-symmetric part, as well as the energy-related mirror-symmetric part, for the homogeneous isotropic two-mode two-time correlation function of a fluctuating field. In \S~\ref{sec:turb_model}, utilizing the  analytical results for the relevant turbulent fluxes obtained by the theoretical formulation, a turbulence model for helical turbulent flows is constructed. In the model, in addition to the eddy-viscosity which contributes to the destruction of large-scale structures, the helicity-related transport coefficient shows up. This helicity effect is expected to contribute to sustainment and generation of large-scale flow structures. In \S~\ref{sec:5}, the helicity turbulence model is applied to a physically interesting and practically important flow configuration: turbulent swirling flow. It will be shown that the helicity model successfully reproduces prominent features of turbulent swirling flow that cannot be reproduced by the standard eddy-viscosity turbulence model. In \S~\ref{sec:6}, global flow generation due to inhomogeneous turbulent helicity is discussed with theoretical and numerical analyses. In \S~\ref{sec:7} we summarize our arguments with conclusion.

\section{Helicity: definition and properties\label{sec:2}}
Helicity possesses several distinctive mathematical and interesting physical properties. After presenting the definition of helicity, these several characteristic properties of helicity will be presented. They include helicity as pseudo-scalar and conserved quantity, geometrical and topological interpretation of helicity, role of helicity in cascade suppression.

\begin{figure}
\centering
\includegraphics{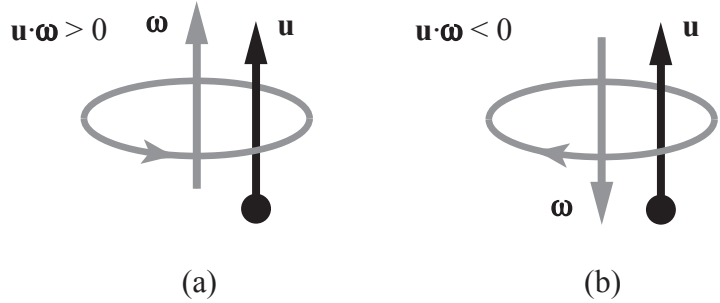}
\caption{
Helicity as the velocity--vorticity correlation. (a) Positive helicity, (b) negative helicity.
\label{fig:1}
}
\end{figure}

\paragraph{(A) Definition}
Helicity is defined as the volume integral of the inner product of the velocity and its curl (vorticity) as
\begin{equation}
	{\cal{H}}
	= \int_V {\bf{u}} \cdot (\nabla \times {\bf{u}})\ dV.	
	\label{eq:vol_int_hel_def}
\end{equation}
Helicity density (hereafter, simply denoted as helicity) is defined by the inner product of the velocity ${\bf{u}}$ and the vorticity $\mbox{\boldmath$\omega$} (= \nabla \times {\bf{u}})$, ${\bf{u}} \cdot \mbox{\boldmath$\omega$}$. A positive helicity (${\bf{u}} \cdot \mbox{\boldmath$\omega$} > 0$) represents a right-handed twist or screw of a fluid element along its motion ${\bf{u}}$. On the other hand, a negative helicity (${\bf{u}} \cdot \mbox{\boldmath$\omega$} < 0$) represents the left-handed twist or screw (Figure~\ref{fig:1}).

It is worth noting that fully helical (Beltrami) situations such as Figure~\ref{fig:1} are unstable. They are fully helical, but eventually become unstable and return to isotropy in the small scales. This means that the constraint of helicity should be less stringent at small scale. Related to this point, the scale dependence of the relative importance of helicity effect in the Reynolds stress will be discussed at the end of \S~\ref{sec:3}.

\paragraph{(B) Pseudo-scalar}
On the basis of parity property, vector quantities are divided into two categories. One is the polar or pure vector, which changes its sign under inversion of coordinate system. The other is the axial or pseudo vector, which does not change its sign under inversion. For the sake of simplicity avoiding complex argument on the transformation between the left- and right-handed system, we just adopt the argument in terms of  ``inversion'' here.\footnote{Transformation of a vector ${\bf{f}} = \{ {f_i} \}$ is written as $f'_i = |{\cal{A}}| {\cal{A}}_{ij} f_j$, where ${\cal{A}}_{ij}$ is the transformation matrix and $|{\cal{A}}|$ is its determinant. For pseudo-vectors $|{\cal{A}}| = -1$. A reflection can be written in terms of combination of inversion ${\cal{I}}$ and rotation ${\cal{C}}$. For instance, a reflection or mirror in the $xy$ plane represented by ${\Sigma} = \left( {\begin{array}{ccc} 1 & 0 & 0\\ 0 & 1 & 0\\ 0 & 0 & -1  \end{array}} \right)$ can be constructed by combination of inversion represented by inversion ${\cal{I}} = \left( {\begin{array}{ccc} -1 & 0 & 0\\ 0 & -1 & 0\\ 0 & 0 & -1 \end{array} } \right)$ and the $\pi$-angle rotation about the $z$ axis ${\cal{C}} = \left( {\begin{array}{ccc} -1 & 0 & 0\\ 0 & -1 & 0\\ 0 & 0 & 1 \end{array}} \right)$ as $\Sigma = {\cal{I}} {\cal{C}}$. Since for all rotation, the determinant $|{\cal{C}}|$ is always $+1$ ($|{\cal{C}}| = +1$), the determinant of the reflection, $|\Sigma|$, is represented by that of the inversion, $|{\cal{I}}|$. In this sense, we can argue non-mirror-symmetry in terms of inversion.}  Under this categorization, the velocity ${\bf{u}}$ is a polar vector while the vorticity $\mbox{\boldmath$\omega$} (= \nabla \times {\bf{u}})$ is an axial vector. It follows that the helicity ${\bf{u}} \cdot \mbox{\boldmath$\omega$}$, defined by the inner product of polar and axial vectors, is a scalar that changes its sign under inversion of coordinate system. Such a scalar is called pseudo-scalar whereas a scalar that does not change its sign under inversion is called pure-scalar. The statistical average of a pseudo-scalar vanishes in mirror-symmetric system. This can be shown as follows. In mirror-symmetric system, by definition of mirror-symmetry, any statistical quantity $f({\bf{x}})$ satisfies $f({\bf{x}}) = f(-{\bf{x}})$ under inversion of coordinate system: ${\bf{x}} \to - {\bf{x}}$. At the same time, by the definition, a pseudo-scalar changes its sign $f({\bf{x}}) = - f(-{\bf{x}})$ under inversion. As this consequence, a pseudo-scalar statistical quantity $f({\bf{x}})$ satisfies $f({\bf{x}}) = - f(-{\bf{x}}) = - f({\bf{x}})$, leading to $f({\bf{x}}) = 0$: The pseudo-scalar statistical quantity vanishes in mirror-symmetric system. Conversely, a finite or non-zero pseudo-scalar implies that the mirror- or reflectional symmetry of the system is broken. Since the helicity is a pseudo-scalar, the helicity serves itself as a measure for broken mirror-symmetry.

\paragraph{(C) Conserved quantity}
One of the prominent characteristics of helicity is that helicity, as well as energy, is an inviscid invariant of the system of fluid dynamics. Namely, the helicity is a conserved quantity in the limit of vanishing viscosity.

	The equations that govern the velocity ${\bf{u}}$ in an incompressible fluid is
\begin{equation}
	\frac{\partial {\bf{u}}}{\partial t}
	+ \left( {{\bf{u}} \cdot \nabla} \right) {\bf{u}}
	= - \nabla p
	+ \nu \nabla^2 {\bf{u}}
	\label{eq:vel_eq}
\end{equation}
or equivalently,
\begin{equation}
	\frac{\partial {\bf{u}}}{\partial t}
	= {\bf{u}} \times \mbox{\boldmath$\omega$}
	- \nabla \left( {p + \frac{{\bf{u}}^2}{2} } \right)
	+ \nu \nabla^2 {\bf{u}},
	\label{eq:vel_eq_rot}
\end{equation}
and the solenoidal condition of ${\bf{u}}$ as
\begin{equation}
	\nabla \cdot {\bf{u}} = 0,
	\label{eq:vel_sol_cond}
\end{equation}
where $p$ is the pressure normalized by a constant density $\rho$, and $\nu$ is the kinematic viscosity. Associated with these equations, the vorticity $\mbox{\boldmath$\omega$} (= \nabla \times {\bf{u}})$ obeys
\begin{equation}
	\frac{\partial \mbox{\boldmath$\omega$}}{\partial t}
	+ \left( {{\bf{u}} \cdot \nabla} \right) \mbox{\boldmath$\omega$}
	= \left( {
    	\mbox{\boldmath$\omega$} \cdot \nabla
	} \right) {\bf{u}}
	+ \nu \nabla^2 \mbox{\boldmath$\omega$}
	\label{eq:vort_eq}
\end{equation}
or equivalently in the rotational form
\begin{equation}
	\frac{\partial \mbox{\boldmath$\omega$}}{\partial t}
	= \nabla \times \left( {
		{\bf{u}} \times \mbox{\boldmath$\omega$}
	} \right)
	+ \nu \nabla^2 \mbox{\boldmath$\omega$},
	\label{eq:vort_eq_rot}
\end{equation}
and the solenoidal condition of $\mbox{\boldmath$\omega$}$ as
\begin{equation}
	\nabla \cdot \mbox{\boldmath$\omega$} = 0.
	\label{eq:sol_vort}
\end{equation}
It follows from Eqs.~(\ref{eq:vel_eq})-(\ref{eq:sol_vort}) that the governing equation of helicity is given by
\begin{eqnarray}
	\frac{d}{dt} {\bf{u}} \cdot \mbox{\boldmath$\omega$}
	&\equiv& \left( {
		\frac{\partial}{\partial t} + {\bf{u}} \cdot \nabla
	} \right) {\bf{u}} \cdot \mbox{\boldmath$\omega$}
	\nonumber\\
	&=& \nabla \cdot \left[ {
    	\left( {\frac{1}{2}{\bf{u}}^2 - p} \right) \mbox{\boldmath$\omega$}
    	+ \nu \nabla ({{\bf{u}} \cdot \mbox{\boldmath$\omega$}})
	} \right]
	- 2 \nu \frac{\partial u_j}{\partial x_i} \frac{\partial \omega_j}{\partial x_i}.
	\label{eq:hel_eq}
\end{eqnarray}
Integrating Eq.~(\ref{eq:hel_eq}) over the volume $V$ and putting $\nu = 0$,\footnote{Of course, putting $\nu = 0$ is not permissible in treating turbulent flows since viscosity plays an essential role in turbulent dissipation. For example, the kinetic helicity decays on a turbulent time scale regardless of the nature of the forcing and the value of the Reynolds number \citep{bra2012}.} we obtain
\begin{equation}
	\frac{d}{dt} \int_V {\bf{u}} \cdot \mbox{\boldmath$\omega$}\ dV
	= \int_S \left[ {
    	\left( {\frac{1}{2}{\bf{u}}^2 - p} \right) \mbox{\boldmath$\omega$}
  	} \right] \cdot {\bf{n}}\ dS.
	\label{eq:cons_tot_hel}
\end{equation}
This means that in the absence of in- or out-flux through the integral surface $S$, the (total) helicity $\int_V {\bf{u}} \cdot \mbox{\boldmath$\omega$}\ dV$ does not vary along the fluid motion ${\bf{u}}$. Namely, the helicity is an inviscid invariant of the fluid motion, as is same as energy (per unit mass) $\int_V {\bf{u}}^2/2\ dV$ is. This fact strongly suggests that helicity as well as energy is a quantity of fundamental importance in fluid dynamics.

\begin{figure}
\centering
\includegraphics{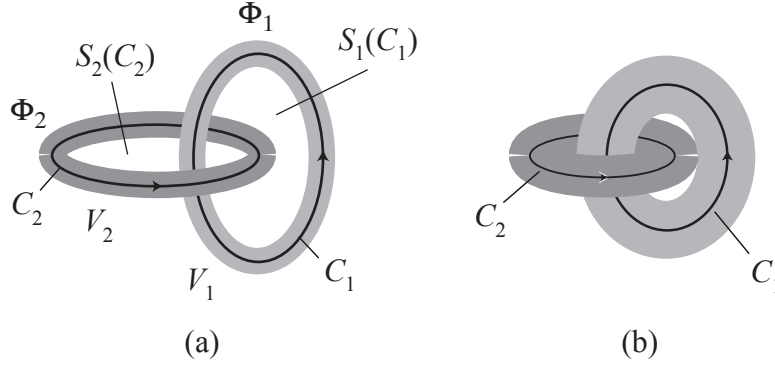}
\caption{
Topological interpretation of helicity. (a) Linked vortex tubes (Hopf link), (b) constraint on relaxation due to helicity invariance.
\label{fig:2}
}
\end{figure}

\paragraph{(D) Geometrical and topological interpretation}
Helicity and its properties can be interpreted and explained in terms of topology \citep{mof1978,mof1992,mof2019}. Helicity corresponds to the topological properties of flows through knots, links, twists, and writhes of the vortex tubes. As a simplest case for links, we consider the Hopf link of two vortex tubes, $\Phi_1$ and $\Phi_2$, which are respectively along the closed loops $C_1$ and $C_2$. They are linked each other as in Figure~\ref{fig:2}(a). The surface areas spanned by the closed loops $C_1$ and $C_2$ are given as $S_1(C_1)$ and $S_2(C_2)$, respectively. The volumes of the vortex-tube regions are denoted as $V_1$ and $V_2$.

	In case that the vorticity is localized only in the regions of $V_1$ and $V_2$, the integral of the helicity over the whole space volume $V$ can be expressed by the sum of the volume integrals over $V_1$ and $V_2$ as
\begin{equation}
	\int_V {\bf{u}} \cdot \mbox{\boldmath$\omega$}\ dV
	= \int_{V_1} {\bf{u}} \cdot \mbox{\boldmath$\omega$}\ dV_1
	+ \int_{V_2} {\bf{u}} \cdot \mbox{\boldmath$\omega$}\ dV_2.
\end{equation}
Assuming that the magnitude of the vorticity $\mbox{\boldmath$\omega$}$ is uniform across the cross section of the vortex-tube region $V_1$, we have
\begin{eqnarray}
	\int_{V_1} {\bf{u}} \cdot \mbox{\boldmath$\omega$}\ dV_1
	&=& \Phi_1 \oint_{C_1} {\bf{u}} \cdot d{\bf{s}}
	= \Phi_1 \int_{S_1(C_1)} \hspace{-20pt} 
		(\nabla \times {\bf{u}}) \cdot {\bf{n}}_1\ dS_1
	\nonumber\\
	&=& \Phi_1 \int_{S_1(C_1)} \hspace{-20pt} 
		\mbox{\boldmath$\omega$}_2 \cdot {\bf{n}}_1\ dS_1
	= \Phi_1 \Phi_2.
\end{eqnarray}
Here use has been made of the assumption that the vorticity is localized so that the vorticity threading its way through the surface $S_1(C_1)$ is solely $\mbox{\boldmath$\omega$}_2$. With a similar calculation on the region $V_2$, the total helicity reads
\begin{equation}
	\int_{V} {\bf{u}} \cdot \mbox{\boldmath$\omega$}\ dV
	= 2 \Phi_1 \Phi_2.
\end{equation}
In general case of the number of links being $N$ ($N$: winding number), the total helicity is expressed as
\begin{equation}
	\int_{V} {\bf{u}} \cdot \mbox{\boldmath$\omega$}\ dV
	= 2N \Phi_1 \Phi_2.
	\label{eq:N_link_num}
\end{equation}
This shows that helicity represents the link of vortex tubes.

	Helicity conservation in the inviscid case corresponds to the fact that the vortex tube is immortal (neither arising nor ceasing) and the linking number of the vortex tubes is invariant in the case of vanishing viscosity ($\nu = 0$).

	Due to the $(\mbox{\boldmath$\omega$} \cdot \nabla) {\bf{u}}$ term in the vorticity equation (\ref{eq:vort_eq}), the vorticity is enhanced (or reduced) by stretching (or shrinking) of the vortex tube. As we saw in (\ref{eq:N_link_num}), helicity conservation is equivalent to the invariance of the linking number. Because of the invariance of the linking number, the vortex shrinking is restricted as in Figure~\ref{fig:2}(b). This topological constraint is linked to the suppression of the energy cascade due to helicity conservation.

\paragraph{(E) Cascade suppression}
Between the velocity and vorticity vectors, ${\bf{u}}$ and $\mbox{\boldmath$\omega$}$, we have the Pythagorean identity as
\begin{equation}
	\frac{({\bf{u}} \cdot \mbox{\boldmath$\omega$})^2}{|{\bf{u}}|^2 |
		\mbox{\boldmath$\omega$}|^2}
	+ \frac{({\bf{u}} \times \mbox{\boldmath$\omega$})^2}{|{\bf{u}}|^2 |
		\mbox{\boldmath$\omega$}|^2}
	= 1.
	\label{eq:pythagorean_id}
\end{equation}
This relationship implies that in a flow with a larger $|{\bf{u}} \cdot \mbox{\boldmath$\omega$}|$, $|{\bf{u}} \times \mbox{\boldmath$\omega$}|$ is relatively smaller. Since ${\bf{u}} \times\mbox{\boldmath$\omega$}$ represents part of the nonlinear interaction in (\ref{eq:vel_eq_rot}) and (\ref{eq:vort_eq_rot}), in a flow with a large $|{\bf{u}} \cdot \mbox{\boldmath$\omega$}|$, the energy cascade due to the nonlinear interaction may be suppressed. With this expectation, the role of helicity in turbulence cascade has been studied for a long time. Through these classical studies, it was shown that the energy dissipation can be locally suppressed in a flow region where the helicity is relatively strong, but with several reservations \citep{rog1987}. It was also suggested that the energy cascade can be suppressed in helical turbulence. However, such a helicity effect on energy-cascade suppression cannot be retained. The effect ceases once the helicity itself starts cascading to smaller scales \citep{and1977}. In order for the helicity effect to be retained in turbulence, helicity has to be sustained by some mechanisms that provide turbulence with helicity. Considering such helicity generation mechanisms, we understand importance of treating the helicity effect in inhomogeneous turbulence such as the case of rotating turbulence, the case with the helicity externally injected into turbulence, etc.

\section{Theoretical analysis of helicity effects in inhomogeneous turbulence\label{sec:3}}
In the previous section, we briefly reviewed how helicity is defined and what kind of properties it possesses. In this section, we present how to address the problem of the helicity effects in inhomogeneous turbulence, and discuss how the turbulent transport might be altered by the presence of helicity. This will be done with the aid of a statistical theoretical analysis of inhomogeneous turbulence.

\subsection{Mean- and fluctuation-fields equations}
We consider a system rotating with a constant angular velocity $\mbox{\boldmath$\omega$}_{\rm{F}}$. We adopt the Reynolds decomposition of a field $f$ as
\begin{equation}
	f = F + f',\;\; F = \langle {f} \rangle
	\label{eq:rey_decomp}
\end{equation}
with 
\begin{subequations}\label{eq:flds_def}
\begin{equation}
	f = ({\bf{u}}, \mbox{\boldmath$\omega$}, p),
	\label{eq:inst_flds}
\end{equation}
\begin{equation}
	F = ({\bf{U}}, \mbox{\boldmath$\Omega$}, P),
	\label{eq:mean_flds}
\end{equation}
\begin{equation}
	f = ({\bf{u}}', \mbox{\boldmath$\omega$}', p'),
	\label{eq:fluct_flds}
\end{equation}
\end{subequations}
where ${\bf{u}}$ is the velocity, $\mbox{\boldmath$\omega$} (= \nabla \times {\bf{u}})$ the vorticity, $p$ the pressure (scaled by a reference density), and $\langle {\cdots} \rangle$ denotes the ensemble averaging. Under this decomposition, the equations governing the mean velocity ${\bf{U}}$ of an incompressible fluid are
\begin{equation}
	\left( {
		\frac{\partial}{\partial t}
		+ {\bf{U}} \cdot \nabla
	} \right) {\bf{U}}
	= {\bf{U}} \times 2 \mbox{\boldmath$\omega$}_{\rm{F}}
	- \nabla P
	- \nabla \cdot \mbox{\boldmath${\cal{R}}$}
	+ \nu \nabla^2 {\bf{U}}
	\label{eq:mean_vel_eq}
\end{equation}
and the solenoidal condition
\begin{equation}
	\nabla \cdot {\bf{U}} = 0,
	\label{eq:mean_vel_sol}
\end{equation}
where $P$ is the mean pressure normalized by the density. The mean vorticity $\mbox{\boldmath$\Omega$} (= \nabla \times {\bf{U}})$ is subject to
\begin{equation}
	\frac{\partial \mbox{\boldmath$\Omega$}}{\partial t}
	= \nabla \times \left[ {
		{\bf{U}} \times (\mbox{\boldmath$\Omega$} 
		+ 2 \mbox{\boldmath$\omega$}_{\rm{F}})
		- \nu \nabla \times \mbox{\boldmath$\Omega$}
	} \right]
	+ \nabla \times {\bf{V}}_{\rm{M}},
	\label{eq:mean_vort_eq}
\end{equation}
\begin{equation}
	\nabla \cdot \mbox{\boldmath$\Omega$} = 0.
	\label{eq:mean_vort_sol}
\end{equation}
In (\ref{eq:mean_vel_eq}) and (\ref{eq:mean_vort_eq}), the Reynolds stress ${\mbox{\boldmath$\cal{R}$}} = \{ {{\cal{R}}_{ij}} \}$ and the ponderomotive or vortexmotive force ${\bf{V}}_{\rm{M}}$ are defined by
\begin{equation}
	{\cal{R}}_{ij}
	= \langle {u'_i u'_j} \rangle,
	\label{eq:rey_strs_def}
\end{equation}
\begin{equation}
	{\bf{V}}_{\rm{M}}
	= \langle {
		{\bf{u}}' \times \mbox{\boldmath$\omega$}'
	} \rangle,
	\label{eq:vmf_def}
\end{equation}
respectively. They represent the sole direct effects of the fluctuations to the mean fields.

	The velocity associated with the rotation with the angular velocity $\mbox{\boldmath$\omega$}_{\rm{F}}$ is ${\bf{u}} = \mbox{\boldmath$\omega$}_{\rm{F}} \times {\bf{x}}$. Taking a curl operation, we have
\begin{eqnarray}
	\mbox{\boldmath$\omega$} 
	&=& \nabla \times {\bf{u}} 
	\nonumber\\
	&=& \nabla \times (\mbox{\boldmath$\omega$}_{\rm{F}} \times {\bf{x}})
	= \mbox{\boldmath$\omega$}_{\rm{F}} (\nabla \cdot {\bf{x}})
	- \mbox{\boldmath$\omega$}_{\rm{F}}
	= 2 \mbox{\boldmath$\omega$}_{\rm{F}}.
	\label{eq:vort_ang_vel_rel}
\end{eqnarray}
This shows that the system rotation with the angular velocity $\mbox{\boldmath$\omega$}_{\rm{F}}$ is locally equivalent to the vorticity of $2\mbox{\boldmath$\omega$}_{\rm{F}}$. We utilize this relationship. Thanks to this local equivalence between the rotation and vorticity, by considering the helicity effect in the turbulence correlations in a rotating frame, we selectively extract the effect of mean vorticity $\mbox{\boldmath$\Omega$} (= \nabla \times {\bf{U}})$ coupled with the helicity using a lower-order calculation.

\subsection{Statistical analytical theory for inhomogeneous turbulence}
In the context of turbulent transport, one of the main aims of statistical analytical theory of turbulence is to obtain the expressions for the turbulent correlations in the mean-field equations. Turbulent correlations, such as the Reynolds stress (\ref{eq:rey_strs_def}) and the vortexmotive force (\ref{eq:vmf_def}), determine the effective transport due to fluctuating motions. The two-scale or multiple-scale direct-interaction approximation (TS- or MS-DIA): a renormalized perturbation expansion theory for strongly nonlinear turbulence combined with the multiple-scale analysis, is a theoretical framework that enables us to treat inhomogeneous turbulence through the derivative expansion. The DIA is one of the oldest modern turbulence theories constituted of renormalized perturbation theory for homogeneous isotropic turbulence at a very large Reynolds number \citep{kra1959}.\footnote{In the DIA formulation, by introducing the response or Green's function as well as the velocity correlation, statistical closure of the system of equations for the propagators (correlation and response functions) is completed. With the aid of the renormalization procedures for the propagators, summations of the expansion terms are performed partially but up to the infinite order of expansion. In terms of the field theory, the DIA is a line (propagator) renormalization theory with truncation of the vertex part at the lowest-order. It is not necessary to invoke this type of elaborate closure scheme for weak nonlinearity turbulence, but according to numerical simulations, the general behavior of DIA on the energy spectrum, energy transfer spectrum, microscale length scale, velocity derivative skewness, etc.\  at low Reynolds numbers was found to be quite good \citep{kra1964}. As for the detailed justification arguments on the DIA, the reader is referred to classical textbooks \citep{les1973,mcc1990}.} As for the notion and detail of the multiple-scale DIA, the reader is referred to \citet{yos1984} and \citet{yok2020}. In this section, we note only some points which should be born in mind in treating the helical turbulence with the aid of the multiple-scale DIA analysis.

\paragraph{(A) Multiple-scale analysis}
The two- or multiple-scale DIA is a combination of the DIA and the multiple-scale analysis. We introduce the slow and fast variables as
\begin{equation}
	{\bf{X}} = \delta_x {\bf{x}},\;\;
	\mbox{\boldmath$\xi$} = {\bf{x}};\;\;
	T = \delta_t t,\;\;
	\tau = t,
	\label{eq:slow_fast_variables}
\end{equation}
where $\delta_x$ and $\delta_t$ are scale parameters of the space and time variables, respectively. If $\delta_x$ and $\delta_t$ are small, ${\bf{X}}$ and $T$ vary significantly only when the original variables ${\bf{x}}$ and $t$ change considerably. In this sense, ${\bf{X}}$ and $T$ are suitable for describing the large-scale and slow evolution, and are called slow variables. If the scale parameters $\delta_x$ and $\delta_t$ are small, we have significant scale separations between the slow and fast variables.

	With these two scale variables ${\bf{X}}$, $\mbox{\boldmath$\xi$}$, $T$, and $\tau$, a field quantity $f({\bf{x}};t)$ is divided into slowly and fast varying components as
\begin{equation}
	f({\bf{x}}; t)
	= F({\bf{X}}; T)
	+ f'({\bf{X}}, \mbox{\boldmath$\xi$}; T, \tau).
	\label{eq:mean_fluct_w_slow_fast_var}
\end{equation}
In the two-scaling formulation, the spatial and temporal derivatives are expressed as
\begin{equation}
	\frac{\partial}{\partial x_i}
	= \frac{\partial}{\partial \xi_i}
	+ \delta_x \frac{\partial}{\partial X_i},\;\;
		\frac{\partial}{\partial t}
	= \frac{\partial}{\partial \tau}
	+ \delta_t \frac{\partial}{\partial T}.
	\label{eq:der_expan}
\end{equation}
This means that the large-scale inhomogeneities appear with the scale parameters $\delta_x$ and $\delta_t$. In this sense, the two-scale analysis is a derivative expansion with respect to the slow-variable inhomogeneities. At this stage, the scale separation affects the validity of the derivative expansion. For instance, in case that the gradient-diffusion-type expression, such as the eddy-viscosity representation, itself is not appropriate for the starting-order approximation, taking only lower-order expansion terms in the the derivative expansion should not give the best approximation.

	We apply this two-scale formulation to the fundamental equations of hydrodynamics. For the simplicity of the analysis, hereafter we assume that the space and time scale parameters are the same and put $\delta_x = \delta_t = \delta$. In the case of an incompressible fluid, the governing equations for the velocity fluctuation are written as
\begin{eqnarray}
	\frac{\partial u'_i}{\partial \tau}
	&+& U_j \frac{\partial u'_i}{\partial \xi_j}
	+ \frac{\partial}{\partial \xi_j} u'_j u'_i
	+ \frac{\partial p'}{\partial \xi_i}
	- \nu \nabla_\xi^2 u'_i
	- 2 \epsilon_{ij\ell} \omega_{{\rm{F}}j} u'_\ell
	\nonumber\\
	&=& \delta \left( {
	- u'_j \frac{\partial U_i}{\partial X_j}
	- \frac{Du'_i}{DT}
	- \frac{\partial p'}{\partial X_i}
	- \frac{\partial}{\partial X_j} \left( {
		u'_j u'_i - \langle {u'_j u'_i} \rangle
	} \right)
	} \right.
	\left. {
	- 2 \nu \frac{\partial^2 u'_i}{\partial X_j \partial \xi_j}
	} \right)
	\nonumber\\
	&+& \delta^2 \left( {\nu \nabla_X^2 u'_i} \right),
	\label{eq:TS_fluct_eq}
\end{eqnarray}
and the solenoidal condition:
\begin{equation}
	\frac{\partial u'_j}{\partial \xi_j}
	+ \delta \frac{\partial u'_j}{\partial X_j}
	= 0,
	\label{eq:TS_sol_cond}
\end{equation}
where the mean-flow advective derivative $D/DT$ is defined as
\begin{equation}
	\frac{D}{DT}
	= \frac{\partial}{\partial T} + {\bf{U}} \cdot \nabla_X.
	\label{eq:mean_adv_der}
\end{equation}

\paragraph{(B) Fourier representation with respect to the fast variables}
We assume that the fluctuation field is homogeneous with respect to the fast space variable $\mbox{\boldmath$\xi$}$. Then a fluctuation field is Fourier transformed as
\begin{equation}
	f'(\mbox{\boldmath$\xi$},{\bf{X}}; \tau,T)
	= \int f'({\bf{k}},{\bf{X}}; \tau,T) 
		\exp[- i {\bf{k}} \cdot (\mbox{\boldmath$\xi$} - {\bf{U}}\tau)]\; 
		d{\bf{k}}.
	\label{eq:TS_fourier_transform}
\end{equation}
Here, the factor ${\bf{k}} \cdot (\mbox{\boldmath$\xi$} - {\bf{U}} \tau)$ denotes that the Fourier transform is performed in the frame traveling with the mean flow ${\bf{U}}$.

	Using this Fourier representation, the equations of the velocity fluctuation in the wave-number space are written as 
\begin{eqnarray}
	\lefteqn{
	\frac{\partial u'_{i}({\bf{k}};\tau)}{\partial \tau}
	+ \nu k^2 u'_{i}({\bf{k}};\tau)
	+ 2 \epsilon_{ij\ell} \omega_{{\rm{F}}j} u'_\ell({\bf{k}};\tau)
	}\nonumber\\
	&-& i k_j \iint \delta({\bf{k}} - {\bf{p}} -{\bf{q}})
		u'_{i}({\bf{p}};\tau) u'_{j}({\bf{q}};\tau)\; d{\bf{p}} d{\bf{q}}
	\nonumber\\
	&=& \delta \left( {
		- u'_j({\bf{k}};\tau) \frac{\partial U_i}{\partial X_j}
		- \frac{Du'_i({\bf{k}};\tau)}{\partial T_{\rm{I}}}
		- \frac{\partial p'}{\partial X_{{\rm{I}}i}}
	} \right.
	\nonumber\\
	&&\left. {
		\hspace{-5pt} - \iint \delta({\bf{k}} - {\bf{p}} - {\bf{q}})
		\frac{\partial}{\partial X_{{\rm{I}}j}} 
		u'_i({\bf{p}};\tau) u'_j({\bf{q}};\tau)\; d{\bf{p}} d{\bf{q}}
		+ \delta({\bf{k}}) \frac{\partial R_{ji}}{\partial X_j}
	} \right),
	\label{eq:TS_vel_eq_fourier}
\end{eqnarray}
and the solenoidal condition:
\begin{equation}
	{\bf{k}} \cdot {\bf{u}}'({\bf{k}};\tau)
	= \delta \left( {
		- i \frac{\partial u'_j({\bf{k}};\tau)}{\partial X_{{\rm{I}}j}}
	} \right),
	\label{eq:TS_sol_cond_fourier}
\end{equation}
where
\begin{equation}
	\left( {
		\frac{\partial}{\partial T_{\rm{I}}}, \nabla_{X{\rm{I}}}
	} \right)
	= \exp(-i {\bf{k}} \cdot {\bf{U}}\tau)
		\left( {\frac{\partial}{\partial T}, \nabla_{X}} \right)
		\exp(i {\bf{k}} \cdot {\bf{U}}\tau).
	\label{eq:interaction_rep}
\end{equation}
The fluctuation fields depend on the slow variables ${\bf{X}}$ and $T$ as well as on the fast variables $\mbox{\boldmath$\xi$}$ and $\tau$. However, for the sake of the simplicity of notation, we suppress denoting ${\bf{X}}$ and $T$ and just denote ${\bf{u}}'(\bf{k}{};\tau)$ in wave-number space.

\paragraph{(C) Scale-parameter expansion}
We expand a field quantity $f$ with the scale parameter $\delta$ as
\begin{equation}
	f' = f_0 + \delta f'_1 + \delta^2 f'_2 + \cdots
	= \sum_n \delta^n f'_n.
	\label{eq:scale_para_expans}
\end{equation}
Substituting (\ref{eq:scale_para_expans}) into the velocity fluctuation equation (\ref{eq:TS_vel_eq_fourier}), with eliminating the pressure $p'$, the lowest or zeroth order velocity equation is given as
\begin{eqnarray}
	\lefteqn{
	\frac{\partial u'_{0i}({\bf{k}};\tau)}{\partial \tau}
	+ \nu k^2 u'_{0i}({\bf{k}};\tau)
	+ 2 \epsilon_{ij\ell} \omega_{{\rm{F}}j} u'_{0\ell}({\bf{k}};\tau)
	}\nonumber\\
	&&\hspace{10pt}- i M_{ij\ell}({\bf{k}}) \iint 
		\delta({\bf{k}} - {\bf{p}} - {\bf{q}}) 
		u'_{0j}({\bf{p}};\tau) u'_{0\ell}({\bf{q}};\tau)\; 
		d{\bf{p}} d{\bf{q}}
	= 0.
	\label{eq:TS_u0_eq}
\end{eqnarray}

\paragraph{(D) External-field expansion}
	As discussed in $\S~2$ helicity is a measure of the reflectional symmetry. In order to delve into the effects of helicity, here we consider hydrodynamic turbulence in a system with rotation, which breaks the reflectional symmetry. The rotation axis gives a particular direction, so the system with rotation is no longer isotropic. In order to utilize some properties of isotropic turbulence, we expand a turbulence field with respect to the angular velocity of rotation $\omega_{\rm{F}} = |\mbox{\boldmath$\omega$}_{\rm{F}}|$. The turbulent velocity field of the $n$-th order in the scale parameter $\delta$ expansion, ${\bf{u}}'_n$ is expanded with $\omega_{\rm{F}}$ as
\begin{equation}
	{\bf{u}}'_n = \sum_{m=0}^\infty \omega_{\rm{F}}^m {\bf{u}}'_{nm}.
	\label{eq:ext_fld_expans}
\end{equation}
Specifically, the lowest-order field ${\bf{u}}_{00}$ is called the basic field and denoted as $ {\bf{u}}'_{\rm{B}}$. The equation for $ {\bf{u}}'_{\rm{B}}$ is written as
\begin{eqnarray}
	\lefteqn{
	\frac{\partial u'_{{\rm{B}}i}({\bf{k}};\tau)}{\partial \tau}
	+ \nu k^2 u'_{{\rm{B}}i}({\bf{k}};\tau)
	}\nonumber\\ 
	&&\hspace{0pt}- i M_{ij\ell}({\bf{k}}) \iint 
		\delta({\bf{k}} - {\bf{p}} - {\bf{q}}) 
		u'_{{\rm{B}}j}({\bf{p}};\tau) u'_{{\rm{B}}\ell}({\bf{q}};\tau)\; 
		d{\bf{p}} d{\bf{q}}
	= 0,
	\label{eq:TS_uB_eq}
\end{eqnarray}
which is exactly the same as that for the homogeneous isotropic turbulence.

	We introduce the Green's function associated with the basic-field equation, $G'_{ij}({\bf{k}};\tau,\tau')$. The equation of $G'_{ij}({\bf{k}};\tau,\tau')$ is defined by
\begin{eqnarray}
	\lefteqn{
	\frac{\partial G'_{ij}({\bf{k}};\tau,\tau')}{\partial \tau}
	+ \nu k^2 G'_{ij}({\bf{k}};\tau,\tau')
	}\nonumber\\
	&&\hspace{0pt}- 2i M_{i\ell m}({\bf{k}}) \iint 
	\delta({\bf{k}} - {\bf{p}} - {\bf{q}}) 
		u'_{{\rm{B}}\ell}({\bf{p}};\tau) G'_{mj}({\bf{q}};\tau,\tau')\; 
		d{\bf{p}} d{\bf{q}}
	\nonumber\\
	&&= D_{ij}({\bf{k}}) \delta(\tau-\tau').
	\label{eq:Gij_eq}
\end{eqnarray}
Reflecting the properties of the basic-field equation (\ref{eq:TS_uB_eq}), the Green's function equation (\ref{eq:Gij_eq}) is the same as the counterpart for the homogeneous isotropic turbulence.

	The first-order equation is given as
\begin{eqnarray}
	\lefteqn{
	\frac{\partial u'_{1i}({\bf{k}};\tau)}{\partial \tau}
	+ \nu k^2 u'_{1i}({\bf{k}};\tau)
	}\nonumber\\
	&&\hspace{0pt}- 2i M_{ij\ell}({\bf{k}}) \iint 
		\delta({\bf{k}} - {\bf{p}} - {\bf{q}}) 
		u'_{{\rm{B}}j}({\bf{p}};\tau) u'_{{\rm{S}}1\ell}({\bf{q}};\tau)\; 
		d{\bf{p}} d{\bf{q}}
	+ 2 \epsilon_{ij\ell} \omega_{{\rm{F}}j} 
		u'_{{\rm{B}}\ell}({\bf{k}};\tau)
	\nonumber\\
	&=& - D_{im}({\bf{k}}) u'_{{\rm{B}}\ell} 
		\frac{\partial U_m}{\partial X_\ell}
	- D_{i\ell}({\bf{k}}) 
		\frac{Du'_{{\rm{B}}\ell}({\bf{k}};\tau)}{DT_{\rm{I}}}
	\nonumber\\
	&&\hspace{0pt}+ 2 M_{i\ell m}({\bf{k}}) \iint 
		\delta({\bf{k}} - {\bf{p}} - {\bf{q}}) 
		\frac{q_m}{q^2} u'_{{\rm{B}}j}({\bf{p}};\tau) 
		\frac{\partial u'_{{\rm{B}}n}}
			{\partial X_{{\rm{I}}n}}({\bf{q}};\tau)\; 
		d{\bf{p}} d{\bf{q}}
	\nonumber\\
	&&\hspace{0pt}+ D_{ip}({\bf{k}}) M_{\ell mnp}({\bf{k}}) 
	\iint \delta({\bf{k}} - {\bf{p}} - {\bf{q}}) 
  		\frac{\partial}{\partial X_{{\rm{I}}n}} 
			u'_{{\rm{B}}\ell}({\bf{p}};\tau) u'_{{\rm{B}}m}({\bf{q}};\tau)\; 
			d{\bf{p}} d{\bf{q}}
	\label{eq:u1_eq}
\end{eqnarray}
with the solenoidal condition based on (\ref{eq:TS_sol_cond}):
\begin{equation}
	{\bf{u}}'_1({\bf{k}};\tau)
	= {\bf{u}}'_{{\rm{S}}1}({\bf{k}};\tau)
	- i \frac{{\bf{k}}}{k^2} 
		\frac{\partial u'_{{\rm{B}}\ell}({\bf{k}};\tau)}
			{\partial X_{{\rm{I}}\ell}},
	\label{eq:u1_sol}
\end{equation}
where ${\bf{u}}'_{{\rm{S}}1}$ is the first-order solenoidal velocity satisfying
\begin{equation}
	\nabla \cdot {\bf{u}}'_{{\rm{S}}1}({\bf{k}};\tau) = 0.
	\label{eq:us1_sol}
\end{equation}

	We see from (\ref{eq:u1_eq}) that all the large-scale inhomogeneity effects enter the right-hand side (r.h.s.) with a scale parameter $\delta$ originated from the r.h.s.\ of (\ref{eq:TS_vel_eq_fourier}). On the other hand, the left-hand side (l.h.s.) of (\ref{eq:u1_eq}) is in the same form as the counterpart in the ${\bf{u}}'_{\rm{B}}$ equation (\ref{eq:TS_uB_eq}). By treating the r.h.s.\ of (\ref{eq:u1_eq}) as the force terms, we formally solve ${\bf{u}}'_{{\rm{S}}1}$ in terms of the Green's function $G$ defined by (\ref{eq:Gij_eq}). Utilizing the expressions of ${\bf{u}}'_{\rm{B}}$, ${\bf{u}}'_{{\rm{S}}1}$, etc., we calculate the turbulent correlations with the aid of the DIA.

\paragraph{(E) Basic field with broken mirror-symmetry}
	In order for the helicity effects to work, the helicity has to be supplied to the turbulence. As will be discussed in \S~4, system rotation coupled with some inhomogeneity provides one of such supply mechanisms. Since the helicity is a measure of the broken mirror-symmetry, we are required to properly capture this breakage of symmetry in the theoretical formulation.
	
	In the standard two-scale analysis, the basic field ${\bf{u}}'_{\rm{B}}$ is assumed to be homogeneous and isotropic. In order to take the helicity effects into account, we assume that the basic field is homogeneous and isotropic but non-mirror-symmetric in this work.
	
\begin{figure}
\centering
\includegraphics[scale=0.8]{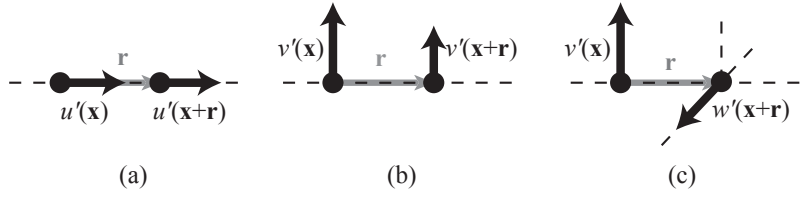}
\caption{
Velocity correlations. (a) Longitudinal, (b) transverse, and (c) cross correlations. Here, ${\bf{r}}$ is the displacement vector, $u'$ is the longitudinal component of the velocity fluctuation along ${\bf{r}}$, $v'$ and $w'$ are the transverse components perpendicular to each other.
\label{fig:3}
}
\end{figure}

	Firstly, we assume a generic expression for the correlation of the basic velocity field in the configuration space. In homogeneous turbulence the statistical properties do not depend on where the origin is located but are determined only by the relative position ${\bf{r}} = {\bf{x}}' - {\bf{x}}$. Then the two-point two-time velocity correlation is written as
\begin{equation}
	\tilde{R}_{ij}({\bf{r}};t,t')
	= \langle {u'_i({\bf{x}};t) u'_j({\bf{x}}';t')} \rangle
	= \langle {u'_i({\bf{0}};t) u'_j({\bf{r}};t')} \rangle.
	\label{eq:homogeneous_correl}
\end{equation}
The general forms of the correlation tensors for various geometries have been derived in \citet{rob1940} and \citet{cha1950}. The generic expression for the isotropic and homogeneous $\tilde{R}_{ij}$ can be obtained as follows \citep{les2008}. From the two-point two-time correlation $\tilde{R}_{ij}$ and two arbitrary fixed vector ${\bf{a}}$ and ${\bf{b}}$, we construct a quantity $a_i \tilde{R}_{ij}({\bf{r}};t,t') b_j$ which is a scalar and consequently isotropic. This scalar quantity should be invariant under rotation of ${\bf{r}}$, ${\bf{a}}$, and ${\bf{b}}$. So, it depends only on the lengths, relative angles, and orientations of this set of vectors; ${\bf{r}} \cdot {\bf{r}}$, ${\bf{a}} \cdot {\bf{a}}$, ${\bf{b}} \cdot {\bf{b}}$, ${\bf{r}} \cdot {\bf{a}}$, ${\bf{r}} \cdot {\bf{b}}$, ${\bf{a}} \cdot {\bf{b}}$, and ${\bf{r}} \cdot ({\bf{a}} \times {\bf{b}})$. Hence, we have
\begin{equation}
	a_i \tilde{R}_{ij}({\bf{r}};t,t') b_j
	= A({\bf{r}};t,t') {\bf{a}} \cdot {\bf{b}}
	+ B({\bf{r}};t,t') ({\bf{r}} \cdot {\bf{a}}) ({\bf{r}} \cdot {\bf{b}})
	+ C({\bf{r}};t,t') {\bf{r}} \cdot ({\bf{a}} \times {\bf{b}}).
	\label{eq:scalar_aRb}
\end{equation}
If we choose ${\bf{a}}$ and ${\bf{b}}$ as the unit vectors in the $i$ and $j$ directions, ${\bf{e}}_i$ and ${\bf{e}}_j$, we obtain
\begin{equation}
	\tilde{R}_{ij}({\bf{r}};t,t')
	= A({\bf{r}};t,t') \delta_{ij}
	+ B({\bf{r}};t,t')r_i r_j
	+ C({\bf{r}};t,t') \epsilon_{ij\ell} r_\ell,
	\label{eq:scalar_R_ABC}
\end{equation}
or equivalently, 
\begin{equation}
	\tilde{R}_{ij}
	= g \left( {\delta_{ij} - \frac{r_i r_j}{r^2}} \right)
	+ f \frac{r_i r_j}{r^2}
	+ h \epsilon_{ij\ell} \frac{r_\ell}{r}
	\label{eq:gene_R_GFH}
\end{equation}
with $g=A$, $f = A + r^2 B$, and $h=rC$. Here $g$ represents the transverse velocity correlation, $f$ the longitudinal velocity correlation, and $h$ is the cross velocity correlation (Figure~\ref{fig:3}). The cross velocity correlation divided by $r$ with limiting $r \to 0$ is equivalent to the local helicity $\langle {{\bf{u}}' \cdot \mbox{\boldmath$\omega$}'} \rangle$. 

Corresponding to (\ref{eq:gene_R_GFH}) in the configuration space, we have an expression in the wave-number space. The statistical property of the basic field is assumed to be
\begin{eqnarray}
	\lefteqn{
	\frac{\langle {
    u'_{{\rm{B}}i}({\bf{k}},{\bf{X}};\tau,T) 
    u'_{{\rm{B}}j}({\bf{k}}',{\bf{X}};\tau',T) 
  } \rangle}{\delta({\bf{k}}+{\bf{k}}')}
	}\nonumber\\
	&&= D_{ij}({\bf{k}}) Q(k,{\bf{X}};\tau,\tau',T)
	+ \Pi_{ij}({\bf{k}}) Q_{\rm{C}}(k,{\bf{X}};\tau,\tau',T)
	\nonumber\\
	&&\hspace{20pt} + \frac{i}{2} \frac{k_\ell}{k^2} \epsilon_{ij\ell} 
		H(k,{\bf{X}};\tau,\tau',T),
	\label{eq:gen_Rij_comp}
\end{eqnarray}
where $Q$ is the energy spectral density of the basic field, $Q_{\rm{C}}$ is the compressible counterpart, and $H$ is the helicity counterpart, $D_{ij}({\bf{k}}) (= \delta_{ij} - k_i k_j/k^2)$ is the solenoidal projection operator, $\Pi_{ij}({\bf{k}}) (= k_i k_j/k^2)$ is the compressible projection operator, $\epsilon_{ij\ell}$ is the alternate tensor. In the case of an incompressible fluid, (\ref{eq:gen_Rij_comp}) is reduced to
\begin{eqnarray}
	\lefteqn{
	\frac{\langle {
    u'_{{\rm{B}}i}({\bf{k}},{\bf{X}};\tau,T) 
    u'_{{\rm{B}}j}({\bf{k}}',{\bf{X}};\tau',T) 
  } \rangle}{\delta({\bf{k}}+{\bf{k}}')}
	}\nonumber\\
	&&= D_{ij}({\bf{k}}) Q(k,{\bf{X}};\tau,\tau',T)
	+ \frac{i}{2} \frac{k_\ell}{k^2} \epsilon_{ij\ell} 
		H(k,{\bf{X}};\tau,\tau',T).
	\label{eq:gen_Rij_sol}
\end{eqnarray}
This is the most generic expression of the homogeneous isotropic non-mirror-symmetric incompressible turbulence \citep{bat1953}. At the same time, the Green's function $G'_{ij}({\bf{k}};\tau,\tau')$ is assumed to be isotropic as
\begin{equation}
	G'_{ij}({\bf{k}};\tau,\tau')
	= D_{ij}({\bf{k}}) G(k;\tau,\tau').
	\label{eq:iso_Gij}
\end{equation}

	As we see (\ref{eq:Rij_config_wave}) in the following subsection (\S~\ref{sec:3.3}), the correlation function in the configuration space can be calculated by the spectral integral of the spectral function. It follows from (\ref{eq:gen_Rij_sol}) that the turbulent energy is calculated as
\begin{equation}
	\langle {{\bf{u}}'_{{\rm{B}}}{}^2} \rangle/2
	= \langle {u'_{{\rm{B}}\ell} u'_{{\rm{B}}\ell}} \rangle/2
	= \int Q(k;\tau,\tau)\ d{\bf{k}},
	\label{eq:energy_Q}
\end{equation}
which shows that the $H$-related part in (\ref{eq:gen_Rij_sol}) does not contribute to the turbulent energy. On the other hand, the turbulent helicity is calculated as
\begin{equation}
	\langle {
		{\bf{u}}'_{{\rm{B}}} \cdot \mbox{\boldmath$\omega$}'_{\rm{B}}
	} \rangle
	= \left\langle {
		u'_{{\rm{B}}\ell} 
		\epsilon_{\ell mn} \frac{\partial u'_{{\rm{B}}n}}{\partial x_m}
	} \right\rangle
	= \int H(k;\tau,\tau)\ d{\bf{k}}.
	\label{eq:helicity_H}
\end{equation}
It is only the non-mirror-symmetric part of (\ref{eq:gen_Rij_sol}) that contributes to the turbulent helicity.

\subsection{Analytical expressions for the Reynolds stress and vortexmotive force\label{sec:3.3}}
The Reynolds stress in the configuration space can be expressed by its wave-number space counterpart as
\begin{equation}
	{\cal{R}}_{ij}
	= \langle {
    	u'_i(\mbox{\boldmath$\xi$},{\bf{X}};\tau,T)
    	u'_j(\mbox{\boldmath$\xi$},{\bf{X}};\tau,T)
	} \rangle
	= \int {\cal{R}}_{ij}({\bf{k}},{\bf{X}};\tau,\tau,T)\ d{\bf{k}},
	\label{eq:Rij_config_wave}
\end{equation}
where the integrand is abbreviated as ${\cal{R}}_{ij}({\bf{k}})$, and is expressed as
\begin{eqnarray}
	{\cal{R}}_{ij}({\bf{k}})
	&=& \frac{\langle {u'_i({\bf{k}};\tau) u'_j({\bf{k}}';\tau)} \rangle}
			{\delta({\bf{k}} + {\bf{k}}')}
	\nonumber\\
	&&\hspace{-30pt}= \frac{\langle {
			u'_{0i}({\bf{k}};\tau) u'_{0j}({\bf{k}}';\tau)
		} \rangle}{\delta({\bf{k}} + {\bf{k}}')}
	\nonumber\\
	&&\hspace{-15pt}+ \delta \left( {
	\frac{\langle {u'_{0i}({\bf{k}};\tau) u'_{1j}({\bf{k}}';\tau)} \rangle}	
	{\delta({\bf{k}} + {\bf{k}}')}
	+ \frac{\langle {u'_{1i}({\bf{k}};\tau) u'_{0j}({\bf{k}}';\tau)}\rangle}
	{\delta({\bf{k}} + {\bf{k}}')}
	} \right)
	+ O(\delta^2)
	\nonumber\\
	&&\hspace{-30pt}= \frac{\langle {
			u'_{{\rm{B}}i}({\bf{k}};\tau) u'_{{\rm{B}}j}({\bf{k}}';\tau)
		} \rangle}{\delta({\bf{k}} + {\bf{k}}')}
	+ \frac{\langle {
		u'_{{\rm{B}}i}({\bf{k}};\tau) u'_{01j}({\bf{k}}';\tau)
	} \rangle}{\delta({\bf{k}} + {\bf{k}}')}
	+ \frac{\langle {
		u'_{01i}({\bf{k}};\tau) u'_{{\rm{B}}j}({\bf{k}}';\tau)
	} \rangle}{\delta({\bf{k}} + {\bf{k}}')}
	\nonumber\\
	&&\hspace{-15pt}+ \delta \left( {
	\frac{\langle {
  		u'_{{\rm{B}}i}({\bf{k}};\tau) u'_{10j}({\bf{k}}';\tau)
	} \rangle}{\delta({\bf{k}} + {\bf{k}}')}
	+ \frac{\langle {
		u'_{10i}({\bf{k}};\tau) u'_{{\rm{B}}j}({\bf{k}}';\tau)
	} \rangle}{\delta({\bf{k}} + {\bf{k}}')}
	+ \cdots
	} \right)
	\nonumber\\
	&&+ O(\delta^2)
	\label{eq:rey_strs_expans}
\end{eqnarray}
Substituting ${\bf{u}}'_{\rm{B}}({\bf{k}};\tau)$ with the statistical properties (\ref{eq:gen_Rij_sol}) and ${\bf{u}}'_{01}({\bf{k}};\tau)$ and ${\bf{u}}'_{10}({\bf{k}};\tau)$ into (\ref{eq:rey_strs_expans}), we obtain the Reynolds stress expression as
\begin{equation}
	{\cal{R}}_{ij}
	= \frac{2}{3} K \delta_{ij}
	- \nu_{\rm{T}} {\cal{S}}_{ij}
	+ \left[ {
			\Gamma_i \Omega_{\ast j} + \Gamma_j \Omega_{\ast i} 
		} \right]_{\rm{D}},
	\label{eq:rey_strs_anal_result}
\end{equation}
where ${\rm{D}}$ denotes the deviatoric or traceless part of a tensor: ${\cal{A}}_{ij{\rm{D}}} = {\cal{A}}_{ij}  - (1/3) {\cal{A}}_{\ell\ell} \delta_{ij}$, $\mbox{\boldmath${\cal{S}}$} = \{ {{\cal{S}}_{ij}} \}$ is the mean velocity strain rate defined by
\begin{equation}
	{\cal{S}}_{ij}
	= \frac{\partial U_i}{\partial x_j}
	+ \frac{\partial U_j}{\partial x_i}
	- \frac{2}{3} \nabla \cdot {\bf{U}} \delta_{ij},
	\label{eq:mean_vel_strain_def}
\end{equation}
and $\mbox{\boldmath$\Omega$}_\ast$ is the mean absolute vorticity $\mbox{\boldmath$\Omega$}_\ast = \mbox{\boldmath$\Omega$} + 2 \mbox{\boldmath$\omega$}_{\rm{F}}$. In (\ref{eq:rey_strs_anal_result}) the transport coefficients $\nu_{\rm{T}}$ and $\mbox{\boldmath$\Gamma$}$ are given as
\begin{equation}
	\nu_{\rm{T}}
	= \frac{7}{15} \int d{\bf{k}} \int_{-\infty}^t \!\!\! d\tau_1
		G(k;\tau,\tau_1) Q(k;\tau,\tau_1),
	\label{eq:nuT_anal_exp}
\end{equation}
\begin{equation}
	\mbox{\boldmath$\Gamma$}
	= \frac{1}{30} \int k^{-2} d{\bf{k}} \int_{-\infty}^t \!\!\! d\tau_1
		G(k;\tau,\tau_1) \nabla H(k;\tau,\tau_1).
	\label{eq:Gamma_anal_exp}
\end{equation}
Here, $\nu_{\rm{T}}$ (\ref{eq:nuT_anal_exp}) is the transport coefficient coupled with the mean velocity strain $\mbox{\boldmath${\cal{S}}$}$ in (\ref{eq:rey_strs_anal_result}), and is called the eddy or turbulent viscosity, which is determined by the Green's function $G$ and the energy spectral function $Q$. In the simplest case where the time integral of the Green's function can be evaluated independent of that of the energy spectral function, $\int_{-\infty}^{t} G(k;\tau,\tau_1) d\tau_1$ gives the turbulence timescale $\tau$, and the spectral integral of energy spectrum $\int d{\bf{k}} Q(k;\tau,\tau_1)$ gives the turbulent energy $K \sim u'{}^2$ ($u'$: magnitude of turbulence velocity). In this case, (\ref{eq:nuT_anal_exp}) is reduced to
\begin{equation}
	\nu_{\rm{T}} \sim \tau K \sim u' \ell,
	\label{eq:nuT_simpl_exp}
\end{equation}
where $\ell \sim u' \tau$ is the characteristic length of turbulence. Equation~(\ref{eq:nuT_simpl_exp}) corresponds to the mixing-length expression for the turbulent viscosity.

	On the other hand, $\mbox{\boldmath$\Gamma$}$ (\ref{eq:Gamma_anal_exp}) is the coupling coefficient for the mean absolute vorticity $\mbox{\boldmath$\Omega$}_\ast$ in (\ref{eq:rey_strs_anal_result}). This $\mbox{\boldmath$\Gamma$}$ is determined by $G$ and the spatial gradient of the turbulent helicity spectral function, $\nabla H$. Equation~(\ref{eq:Gamma_anal_exp}) implies that the inhomogeneity of the turbulent helicity is essential for the helicity effect in the momentum transport. In a similar manner as in (\ref{eq:nuT_simpl_exp}), the simplest expression for $\mbox{\boldmath$\Gamma$}$ can be written as
\begin{equation}
	\mbox{\boldmath$\Gamma$} \sim \ell^2 \tau \nabla H.
	\label{eq:Gamma_simpl_exp}
\end{equation}
This form of $\mbox{\boldmath$\Gamma$}$; proportional to $\nabla H$ instead of $H$, can be naturally understood by symmetry argument of the Reynolds stress. The Reynolds stress $\mbox{\boldmath$\cal{R}$} = \{ {\langle {u'_i u'_j} \rangle} \}$ is symmetric: ${\cal{R}}_{ij} \to \hat{\cal{R}}_{ij} = \langle {\hat{u}'_i \hat{u}'_j} \rangle = \langle {(-u'_i) (- u'_j)} \rangle = \langle {u'_i u'_j} \rangle = {\cal{R}}_{ij}$ with respect to inversion of the coordinate system ${\bf{x}} \to \hat{\bf{x}} = - {\bf{x}}$ and has even parity. The mean (absolute) vorticity is an axial- or pseudo-vector which does not change its sign under inversion. On the other hand, the turbulent helicity is a pseudo-scalar which changes its sign $H \to \hat{H} = -H$ under inversion. As this consequence, the helicity $H$ (odd parity) itself cannot enter the Reynolds-stress expression as the proportional coefficient coupled with $\mbox{\boldmath$\Omega$}_\ast$ but the gradient of helicity, $\nabla H$, which has even parity $\nabla H \to \hat{\nabla} \hat{H} = (-\nabla) (-H) = \nabla H$, can. Actually, it is this $\nabla H$ dependence that was obtained in (\ref{eq:Gamma_anal_exp}). 

Note that $\mbox{\boldmath$\Gamma$}$ is a vector and that the direction of $\mbox{\boldmath$\Gamma$}$ is determined by the spatial distribution of the turbulent helicity $H$ through $\nabla H$. Depending on the spatial distribution of $H$, the sign of each component of $\Gamma_i \Omega_{\ast j} + \Gamma_j \Omega_{\ast i}$ in (\ref{eq:rey_strs_anal_result}) can be positive or negative. If its sign is in the opposite sense as the eddy-viscosity term $- \nu_{\rm{T}} {\cal{S}}_{ij}$, the $\mbox{\boldmath$\Gamma$}$- or helicity-related term suppresses the eddy-viscosity effect. Otherwise, it enhances the eddy-viscosity effect. However, the sign of the $\mbox{\boldmath$\Gamma$}$- or helicity-related term is not arbitrarily distributed. Since the generation of turbulent helicity depends on the mean vorticity and rotation, it is often expected that the helicity inhomogeneity coupled with the mean vorticity and rotation, $\mbox{\boldmath$\Gamma$} \mbox{\boldmath$\Omega$}_\ast = \{ {\Gamma_i \Omega_{\ast j}} \}$, shows a certain sign. Detailed discussions should be done on the basis of the transport equation of turbulent helicity ((\ref{eq:F_eq}) with (\ref{eq:PH_def})-(\ref{eq:TH_def}) and its model (\ref{eq:H_model_eq}) in \S~\ref{sec:hel_turb_model}).

	The vortexmotive force ${\bf{V}}_{\rm{M}} = \{ {V_{{\rm{M}}i}} \}$ in the mean vorticity equation (\ref{eq:mean_vort_eq}) can be obtained from the Reynolds stress $\mbox{\boldmath$\cal{R}$} = \{ {{\cal{R}}_{ij}} \}$ through the exact relationship:
\begin{equation}
	{\bf{V}}_{\rm{M}} 
	= - \nabla \cdot \mbox{\boldmath$\cal{R}$}
	+ \nabla K,
	\label{eq:VM_rey_strs_rel}
\end{equation}
where $K (= \langle {{\bf{u}}'{}^2} \rangle/2)$ is the turbulent energy. The second or $K$-related term does not contribute to the evolution of the mean vorticity $\mbox{\boldmath$\Omega$}$ since $\nabla \times (\nabla K) = 0$.

	Substituting the Reynolds-stress expression (\ref{eq:rey_strs_anal_result}) into (\ref{eq:VM_rey_strs_rel}), we have the expression of the vortexmotive force as
\begin{equation}
	{\bf{V}}_{\rm{M}}
	= - D_\Gamma (\mbox{\boldmath$\Omega$} 
	+ 2\mbox{\boldmath$\omega$}_{\rm{F}}) 
	- [ ({\mbox{\boldmath$\Omega$} 
	+ 2\mbox{\boldmath$\omega$}_{\rm{F}}}) \cdot \nabla]
    	\mbox{\boldmath$\Gamma$}
	- \nu_{\rm{T}} \nabla \times \mbox{\boldmath$\Omega$}
	\label{eq:vmf_anal_result}
\end{equation}
with
\begin{equation}
	D_\Gamma = \nabla \cdot \mbox{\boldmath$\Gamma$}.
	\label{eq:DGamma_def}
\end{equation}
In (\ref{eq:vmf_anal_result}), the third or $\nu_{\rm{T}}$-related term represents the destruction of the mean vorticity $\mbox{\boldmath$\Omega$}$ due to the eddy viscosity. On the other hand, the first and second terms related to $\mbox{\boldmath$\Gamma$}$ represent the possible generation or sustainment of $\mbox{\boldmath$\Omega$}$ against the eddy-viscosity effect. We see from (\ref{eq:vmf_anal_result}) that the inhomogeneous turbulent helicity coupled with the mean absolute vorticity (mean vorticity and rotation) may contribute to the mean-vorticity generation whereas the turbulent energy does to the mean-vorticity destruction.

	From the spectral expression of the Reynolds stress (\ref{eq:rey_strs_anal_result}) with (\ref{eq:nuT_anal_exp}) and (\ref{eq:Gamma_anal_exp}), the relative magnitude of the helicity effect to the eddy-viscosity effect may be written as
\begin{equation}
	\frac{\mbox{(helicity effect)}}{\mbox{(eddy-viscosity effect)}}
	= \frac{|\nu_{\rm{T}} \mbox{\boldmath${\cal{S}}$}|}
		{|\mbox{\boldmath$\Gamma$} \mbox{\boldmath$\Omega$}_\ast|}
	\sim \frac{\Omega_\ast}{{\cal{S}}}
		\frac{|H(k)|}{kE(k)},
	\label{eq:rel_relevance_hel_effect}
\end{equation}
where $\Omega_\ast = \sqrt{\Omega_{\ast ij} \Omega_{\ast ij}}$ and ${\cal{S}} = \sqrt{{\cal{S}}_{ij} {\cal{S}}_{ij}}$ are the magnitudes of the mean absolute vorticity and the mean velocity strain, respectively. If the scaling of the helicity spectrum $H(k)$ is the same as that of the energy $E(k)$, which is supported by DNSs, the relative helicity $|H(k)|/[kE(k)]$ behaves as $\propto k^{-1}$. This implies that the relative importance of the helicity effect decreases as we go to the smaller scale. This point should be born in mind when we consider the application of the helicity effect to subgrid-scale (SGS) modeling of large-eddy simulations (LESs) with a small filter width $\Delta$.

\section{Turbulence modeling with helicity\label{sec:turb_model}}
	In the previous sections, we showed that the mean velocity and vorticity fields are subjected to the effects of turbulent helicity through the Reynolds stress $\mbox{\boldmath${\cal{R}}$} = \{ {{\cal{R}}_{ij}} \}$ and the turbulent vortexmotive force ${\bf{V}}_{\rm{M}} = \{ {V_{{\rm{M}}i}} \}$. From the practical viewpoint, it is important to construct a turbulence model with incorporating the helicity effects. Helical turbulence, in which the velocity and vorticity fluctuations are (at least partly) aligned with each other, is essentially three dimensional. At the same time, the helical flow structure often shows a strong anisotropy at large scales. As a consequence, we have to consider a three-dimensional huge simulation domain throughout all scales ranging from large to small scales. This is a very demanding situation for direct numerical simulations (DNSs). With the aid of a turbulence model incorporating the helicity effect, we can consistently perform numerical analyses of the helical turbulent flow configuration relevant to astro-, geo-physical, plasma physics, engineering phenomena. Here, we present an example of such a helicity turbulence model with emphasis on the basic notion of the modeling.

\subsection{Choice of one-point statistical quantities}
	As was pointed out in \S~2, the total amount of helicity, as well as that of energy, is an inviscid invariant of the system of fluid equations. In this sense, the local density of helicity, as well as the local density of energy, is a fundamental quantity that represents the properties of fluid dynamics.

	The turbulent energy $\langle {{\bf{u}}'{}^2} \rangle/2 (\equiv K)$ represents the local intensity of fluctuating motions. This point is well reflected by the fact that the turbulent energy determines the turbulent transport such as the eddy viscosity. In contrast to the turbulent energy, the turbulent helicity $\langle {{\bf{u}}' \cdot \mbox{\boldmath$\omega$}'} \rangle (\equiv H)$ represents the left- and right-handed twist property. In this sense, the turbulent helicity can serve itself as a measure of the structure of fluctuating motions.

	The $K-\varepsilon$ model is one of the most used turbulence models in engineering and scientific fields. In the model, in order to solve both the mean and turbulent fields in a self-consistent manner, equations of the turbulent statistical quantities, the turbulent energy $K = \langle {{\bf{u}}'{}^2} \rangle/2$ and its dissipation rate $\varepsilon$, are simultaneously solved with the mean-field equations \citep{lau1972,rod1993}. The turbulent transport coefficients, such as the eddy viscosity, eddy diffusivity, etc., are expressed in terms of $K$ and $\varepsilon$. In the sense that the transport equations are solved with the dynamics of the mean fields, the $K-\varepsilon$ model is much more elaborated in nonlinear and self-consistent treatments of the mean--turbulence interaction.

	In addition to the turbulent energy per mass $K$ and its dissipation rate $\varepsilon$, here we adopt the turbulent helicity $H$ as one of the turbulent statistical quantities. These three turbulent statistical quantities are defined as
\begin{equation}
	K = \langle {{\bf{u}}'{}^2} \rangle/2,
	\label{eq:K_def}
\end{equation}
\begin{equation}
	\varepsilon 
	= \nu \left\langle {\frac{\partial u'_j}{\partial x_i} 
		\frac{\partial u'_j}{\partial x_i}} \right\rangle,
	\label{eq:eps_def}
\end{equation}
\begin{equation}
	H = \langle {{\bf{u}}' \cdot \mbox{\boldmath$\omega$}'} \rangle.
	\label{eq:H_def}
\end{equation}
In addition, we can adopt the dissipation rate of turbulent helicity, $\varepsilon_H$:
\begin{equation}
	\varepsilon_H
	= 2\nu \left\langle {
		\frac{\partial u'_j}{\partial x_i}
		\frac{\partial \omega'_j}{\partial x_i} 
	} \right\rangle
	\label{eq:eps_H_def}
\end{equation}
as a turbulent statistical quantity. However, as compared with the model equation of $\varepsilon$, the counterpart of $\varepsilon_H$ has not been well formulated. So, we do not adopt $\varepsilon_H$ as a turbulent statistical quantities. As for the modeling of helicity dissipation-rate equation, see \citet{yok2016a}.

	The expressions of the eddy viscosity $\nu_{\rm{T}}$ (\ref{eq:nuT_anal_exp}) and the helicity-related transport coefficient $\mbox{\boldmath$\Gamma$}$ (\ref{eq:Gamma_anal_exp}) with the integrals of the spectral and response functions in the wave-number space are too heavy for a practical use. Using the above one-point statistical quantities, $K$, $\varepsilon$, and $H$, the eddy viscosity $\nu_{\rm{T}}$ and the helicity-related coefficient $\mbox{\boldmath$\Gamma$}$ in the Reynolds stress $\mbox{\boldmath${\cal{R}}$}$ (\ref{eq:rey_strs_anal_result}) and the vortexmotive force ${\bf{V}}_{\rm{M}}$ (\ref{eq:vmf_anal_result}) are simplified and modeled as
\begin{equation}
	\nu_{\rm{T}} = C_\mu \frac{K^2}{\varepsilon},
	\label{eq:nuT_model_Keps}
\end{equation}
\begin{equation}
	\mbox{\boldmath$\Gamma$} 
	= C_\gamma \frac{K^4}{\varepsilon^3} \nabla H,
	\label{eq:Gamma_model_KepsH}
\end{equation}
where $C_\mu $ and $C_\gamma$ are model constants. Note that these expressions are written in terms of the turbulent statistical quantities $K$, $\varepsilon$, and $H$ with $\tau \sim K/\varepsilon$ and $\ell \sim K^{3/2}/\varepsilon$, while the expressions (\ref{eq:nuT_simpl_exp}) and (\ref{eq:Gamma_simpl_exp}) are expressed in terms of the mixing length $\ell$ and the characteristic timescale $\tau$ of the turbulence.

\subsection{Helicity turbulence model\label{sec:hel_turb_model}}
The helicity turbulence model is constituted of the mean-velocity equation (\ref{eq:mean_vel_eq}) with the Reynolds stress expression (\ref{eq:rey_strs_anal_result}) accompanied by the transport coefficients $\nu_{\rm{T}}$ (\ref{eq:nuT_anal_exp}) and $\mbox{\boldmath$\Gamma$}$ (\ref{eq:Gamma_anal_exp}), and the model equations for the three turbulent statistical quantities, $K$, $\varepsilon$, and $H$.

	From the fundamental equations, the exact equations of the turbulent energy and helicity are given as
\begin{equation}
	\left( {
		\frac{\partial}{\partial t} 
		+ {\bf{U}} \cdot \nabla
	} \right) F
	= P_F
	- \varepsilon_F
	+ \nabla \cdot {\bf{T}}_F,
	\label{eq:F_eq}
\end{equation}
where $F$ denotes the turbulent energy $K$ or turbulent helicity $H$ as $F = (K, H)$. In Eq.~(\ref{eq:F_eq}), $P_F$, $\varepsilon_F$, and ${\bf{T}}_F$ are the production rate, dissipation rate, and transport rate flux of $F$, respectively. They are defined by
\begin{equation}
	P_K
	= - \left\langle {u'_\ell u'_m} \right\rangle 
	\frac{\partial U_m}{\partial x_\ell},
	\label{eq:PK_def}
\end{equation}
\begin{equation}
	\varepsilon_K
	= \nu \left\langle {
	\frac{\partial u'_m}{\partial x_\ell} 
	\frac{\partial u'_m}{\partial x_\ell}
	} \right\rangle
	\equiv \varepsilon,
	\label{eq:epsK_eq}
\end{equation}
\begin{equation}
	{\bf{T}}_K
	= - \left\langle {p' {\bf{u}}'} \right\rangle
	- \left\langle {\frac{1}{2} {\bf{u}}'{}^2 {\bf{u}}'} \right\rangle
	+ \nu \nabla \left( {
	\frac{1}{2} \left\langle {{\bf{u}}'{}^2} \right\rangle
	} \right),
	\label{eq:TK_def}
\end{equation}
\begin{equation}
	P_H
	= - \left\langle {
		u'_m u'_\ell
	} \right\rangle \frac{\partial \Omega_m}{\partial x_\ell}
	+ \Omega_\ell \frac{\partial}{\partial x_m} \left\langle {
		u'_m u'_\ell
	} \right\rangle,
	\label{eq:PH_def}
\end{equation}
\begin{equation}
	\varepsilon_H
	= 2 \nu \left\langle {
		\frac{\partial u'_m}{\partial x_\ell} 
		\frac{\partial \omega'_m}{\partial x_\ell}
	} \right\rangle,
	\label{eq:epsH_def}
\end{equation}
\begin{equation}
	{\bf{T}}_H
	= -\left\langle {
		p' \mbox{\boldmath$\omega$}'
	} \right\rangle 
	+\frac{1}{2} \left\langle {
		{\bf{u}}'{}^2 \mbox{\boldmath$\omega$}'
	} \right\rangle
	- \left\langle {
		({\bf{u}}' \cdot \mbox{\boldmath$\omega$}') {\bf{u}}'
	} \right\rangle
	+ 2 \mbox{\boldmath$\omega$}_{\rm{F}} \cdot 
		\langle {{\bf{u}}' {\bf{u}}'} \rangle
	+ \nu \nabla \left\langle {
		{\bf{u}}' \cdot \mbox{\boldmath$\omega$}'
	} \right\rangle,
	\label{eq:TH_def}
\end{equation}
where $[2 \mbox{\boldmath$\omega$}_{\rm{F}} \cdot \langle {{\bf{u}}' {\bf{u}}'} \rangle]_\ell = 2\omega_{{\rm{F}}m} \left\langle {u'_m u'_\ell} \right\rangle$.

	From the exact equation (\ref{eq:F_eq}) with (\ref{eq:PK_def})-(\ref{eq:TH_def}), the model equations of $K$, $\varepsilon$, and $H$ may be given as
\begin{equation}
	\left( {
		\frac{\partial}{\partial t}
		+ {\bf{U}} \cdot \nabla
	} \right) K
	= P_K
	- \varepsilon
	+ \nabla \cdot \left( {
		\frac{\nu_{\rm{T}}}{\sigma_K} \nabla K
	} \right),
	\label{eq:K_model_eq}
\end{equation}
\begin{equation}
	\left( {
		\frac{\partial}{\partial t}
		+ {\bf{U}} \cdot \nabla
	} \right) \varepsilon
	= C_{\varepsilon 1} \frac{\varepsilon}{K} P_K
	- C_{\varepsilon 2} \frac{\varepsilon}{K} \varepsilon
	+ \nabla \cdot \left( {
		\frac{\nu_{\rm{T}}}{\sigma_\varepsilon} \nabla \varepsilon
	} \right),
	\label{eq:eps_model_eq}
\end{equation}
\begin{equation}
	\left( {
		\frac{\partial}{\partial t}
	+ {\bf{U}} \cdot \nabla
	} \right) H
	= P_H
	- \varepsilon_H
	+ \nabla \cdot \left( {
    	2 \mbox{\boldmath$\omega$}_{\rm{F}} K
    + \frac{\nu_{\rm{T}}}{\sigma_H} \nabla H
	} \right).
	\label{eq:H_model_eq}
\end{equation}
Here, $\sigma_K$, $\sigma_\varepsilon$, $\sigma_H$ are the turbulent Prandtl numbers for the energy, dissipation rate, and helicity, $C_{\varepsilon 1}$, $C_{\varepsilon_2}$ and $C_H$ are the model constants related to the energy and helicity dissipation rates. 

	For $\varepsilon_H$ in the r.h.s.\ of (\ref{eq:H_model_eq}), we adopt an algebraic model as 
\begin{equation}
	\varepsilon_H 
	= C_H \frac{\varepsilon}{K} H.
	\label{eq:eps_H_alg_model}
\end{equation}

	Alternatively, we can construct the transport equation of the helicity dissipation rate $\varepsilon_H$ and solve it as well as (\ref{eq:K_model_eq})-(\ref{eq:H_model_eq}) without resorting to the algebraic model (\ref{eq:eps_H_alg_model}). We assume that, like a passive scalar, the helicity is determined by the scale $k$, energy and helicity transfer rates, $\varepsilon$ and $\varepsilon_H$. The spectra in the inertial range are assumed to be
\begin{equation}
	\frac{\sigma_K(k,{\bf{x};t})}{\varepsilon} 
	= \sigma_{K0} \varepsilon^{-1/3} k^{-11/3},
	\label{eq:en_spectrum}
\end{equation}
\begin{equation}
	\frac{\sigma_H(k,{\bf{x};t})}{\varepsilon_H} 
	= \sigma_{H0} \varepsilon^{-1/3} k^{-11/3},
	\label{eq:hel_spectrum}
\end{equation} 
where $\sigma_{K0}$ is the Kolmogorov constant and $\sigma_{H0}$ is the counterpart for helicity. With the aid of TS-DIA, up to the lowest-order analysis, we obtain the algebraic relation among $H$, $\varepsilon$, $\varepsilon_H$, and the size of the largest energy-containing eddies $\ell_{\rm{C}}$ as (\ref{eq:eps_H_alg_model}). This corresponds to the estimate of the turbulent helicity dissipation rate $\varepsilon_H$ in HIT. If we proceed to the first-order analysis, the equation of the helicity dissipation rate $\varepsilon_H$ is obtained as
\begin{equation}
	\frac{D\varepsilon_H}{Dt}
	= C_{\varepsilon H 1} \frac{\varepsilon_H}{K} P_K
	- C_{\varepsilon H 2} \frac{\varepsilon_H}{K} \varepsilon
	+ C_{\varepsilon H 3} \frac{\varepsilon_H}{H} P_H
	- C_{\varepsilon H 4} \frac{\varepsilon_H}{H} \varepsilon_H,
	\label{eq:eps_H_eq}
\end{equation}
where the model constants are theoretically estimated as
\begin{equation}
	C_{\varepsilon H 1}=0.36,\;\; 
	C_{\varepsilon H 2}=0.49,\;\;
	C_{\varepsilon H 3}=C_{\varepsilon H 4}=1.1.
	\label{eq:eps_H_model_consts}
\end{equation}
Reflecting the assumption that the helicity spectrum depends both $\varepsilon$ and $\varepsilon_H$ as (\ref{eq:hel_spectrum}), the $\varepsilon_H$ equation depends not only $P_H$ and $\varepsilon_H$ but also on $P_K$ and $\varepsilon$. For the details of the derivation of the model, see \citet{yok2016a}. 

	Since the validity of the $\varepsilon_H$ equation has not been numerically examined yet, hereafter we just adopt (\ref{eq:H_model_eq}) with the algebraic model (\ref{eq:eps_H_alg_model}).\footnote{However, it should be noted that, in some flow configurations, the algebraic model results in discrepancy. For instance, in DNSs of the Ekman boundary layer, the spatial distribution of the turbulent helicity $H$ shows a sign reversal at some hight, while the dissipation rate of the turbulent helicity $\varepsilon_H$ does not vanish there \citep{deu2014}.}

	In the equations of the mean-flow energy and helicity, ${\bf{U}}^2/2$ and ${\bf{U}} \cdot \mbox{\boldmath$\Omega$}$, we have exactly the same terms but the opposite signs to $P_K$ (\ref{eq:PK_def}) and $P_H$ (\ref{eq:PH_def}). This means that the sink (or source) of the mean-field energy and helicity work as the source (or sink) of the turbulent counterparts. In this sense, the production terms $P_K$ and $P_H$ represent the cascades from the mean-field energy and helicity to the turbulent counterparts. We see from (\ref{eq:PK_def}) and (\ref{eq:PH_def}) that inhomogeneous mean fields coupled with the Reynolds stress cause such cascades and contribute to production of the turbulent energy and helicity.

	As we see in the third or $\sigma$-related terms of the model equations (\ref{eq:K_model_eq})-(\ref{eq:H_model_eq}), the transport rate terms such as ${\bf{T}}_K$, ${\bf{T}}_H$, etc.\ are modeled with the gradient diffusion approximations. On the other hand, the $2 \mbox{\boldmath$\omega$}_{\rm{F}} K$ term in (\ref{eq:H_model_eq}) represents the turbulent helicity generation arising from the inhomogeneity along the rotation:
\begin{equation}
	\nabla \cdot \left( {
		2 \mbox{\boldmath$\omega$}_{\rm{F}} K
	} \right)
	= (2 \mbox{\boldmath$\omega$}_{\rm{F}} \cdot \nabla) K.
	\label{eq:inhomo_K_along_rot}
\end{equation}
This term is expected to play a crucial role in turbulent helicity generation in geophysical and astrophysical flows, where rotation and stratification are key ingredients. For instance, in the case of typhoon and hurricane, the rotation $\mbox{\boldmath$\omega$}_{\rm{F}}$ coupled with the vertical inhomogeneity of flow may contribute to the turbulent helicity generation. By contrast, in the case of tornado, the system rotation $\mbox{\boldmath$\omega$}_{\rm{F}}$ will not work, but the mean relative vorticity coupled with the inhomogeneous Reynolds stress in the second term of $P_H$ (\ref{eq:PH_def}) may contribute to the generation of turbulent helicity. Possible applications to some geo- and astro-physical flows will be discussed in \S~\ref{sec:6.4}

	In the absence of the turbulent helicity ($H=0$), this three-equation ($K-\varepsilon-H$) model should be reduced to the standard two-equation ($K-\varepsilon$) model. So, the model constant related to the $K-\varepsilon$ model should be retained the same as the ones in the standard $K-\varepsilon$ model as
\begin{equation}
	C_\mu = 0.09,\; \sigma_K = 1.0,\; C_{\varepsilon 1} = 1.4,\; 	
	C_{\varepsilon 2} = 1.9,\; \sigma_\varepsilon = 1.3.
	\label{eq:model_consts}
\end{equation}
Note that the $K-\varepsilon$ model with the fixed constants has been successfully applied to many two- and three-dimensional wall boundary layers, duct flows, free shear flows, recirculating flows, confined flows, and jets. A useful account of model and model constants can be seen in \citet{lau1972,rod1993}. In this sense, the model constants (\ref{eq:model_consts}) in the $K-\varepsilon$ model has been well optimized and should be fixed as far as possible. It is known that a 5\% change in $C_{\varepsilon 1}$ or $C_{\varepsilon 2}$ results in 20\% change of the spreading rate of a jet \citep{rod1993}. As will be discussed later in \S~\ref{sec:5.3}, without the helicity effect, a considerable reduction of $C_\mu = 0.09 \to 0.001$ is required for a practical application of the standard $K-\varepsilon$ model to a swirling pipe flow. However, there is no firm basis for this ad hoc treatment.

	For the rest of the model constants which are intrinsic to the $K-\varepsilon-H$ model, they should be determined through the practical applications of the present helicity model to helical turbulent flows such as swirling flows.

	Utilizing the notion of the helicity effect, we can construct a set of SGS turbulence model for large-eddy simulations (LESs). In the LESs, we apply a filter to a field quantity $f$ to divide it into the grid-scale (GS) and subgrid-scale (SGS) components, $\overline{f}$ and $f''$. In the simulation, the large-scale or GS motions (large eddies) are solved, whereas the small-scale or SGS motions are modeled (SGS modeling). The Smagorinsky model is the most commonly used SGS model. In the model, the SGS viscosity $\nu_{\rm{S}}$ appearing in the GS equation is expressed in terms of the GS velocity strain $\overline{S} (= \sqrt{{\overline{\cal{S}}}_{ij}^2/2})$ and the filter width $\Delta$ as $\nu_{\rm{S}} = (C_{\rm{S}} \Delta)^2 \overline{S}$ ($\overline{\cal{S}}_{ij}$: GS rate-of-strain tensor, $C_{\rm{S}}$: the Smagorinsky constant). As this expression implies, the Smagorinsky model is the SGS counterpart of the mixing-length model (Note that $\Delta$ is the largest scale of the unresolved or SGS motions.). This simplicity has made the Smagorinsky model be commonly used in various types of turbulent flows. However, there are several deficiencies in the Smagorinsky model. One of such deficiencies is need for constant adjustment. The Smagorinsky constant $C_{\rm{S}}$ has to be adjusted from flow to flow: $C_{\rm{S}} = 0.18$ for the isotropic turbulence, $0.15$ for mixing layer turbulence, $0.10$ for wall turbulence. This means that the Smagorinsky model with model constant $C_{\rm{S}} = 1.8$, optimized for the homogeneous isotropic turbulence, results in more than three times ($(1.8/1.0)^2 = 3.24$) too dissipative if it is applied to the near wall turbulence where the streamwise vorticity structures are ubiquitously observed.  
	
	Since the streamwise vorticity strongly implies the local presence of turbulent helicity, it is expected that the turbulent helicity plays an important role for reducing the effective viscosity in shear turbulence such as the mixing layer and wall turbulence. In the model, the effects of turbulence structure represented by the streamwise vorticity are incorporated into the SGS model through the SGS helicity. The problem of the overestimate of the SGS viscosity in the presence of the subgrid-scale (SGS) shear structure including the streamwise vorticity is alleviated by the implementation of the SGS helicity effect even in the framework of classical Smagorinsky model \citep{yok2017}.

\section{Application to swirling flow\label{sec:5}}

\subsection{Swirling flows}
The helicity turbulence model is constituted of the mean velocity equation and the turbulent field equations, where the Reynolds stress is given by (\ref{eq:rey_strs_anal_result}) with the turbulent viscosity $\nu_{\rm{T}}$ and helicity-related coefficient $\mbox{\boldmath$\Gamma$}$ expressed as (\ref{eq:nuT_model_Keps}) and (\ref{eq:Gamma_model_KepsH}), respectively. The model equations of the turbulent statistical quantities are (\ref{eq:K_model_eq}) for the turbulent energy $K$, (\ref{eq:eps_model_eq}) for the energy dissipation rate $\varepsilon$, and (\ref{eq:H_model_eq}) for the turbulent helicity $H$ with an algebraic model for the helicity dissipation rate $\varepsilon_H$ (\ref{eq:eps_H_alg_model}).

\begin{figure}
\centering
\includegraphics[scale=1.2]{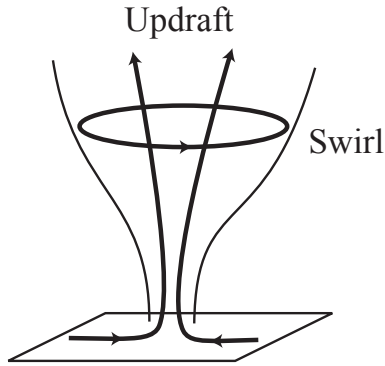}
\caption{
Cyclone swirl accompanied by updraft.
\label{fig:4}
}
\end{figure}

	We apply the present helicity model to turbulent swirling flows. Firstly, we summarize the characteristic properties of swirling flow. Swirling flow consists of the main or axial flow in the mainstream or axial direction and the swirl or circumferential flow around the axis. This configuration is so simple that the swirling flow is ubiquitously observed in the scientific and engineering communities. In the engineering field,  swirling flow is often utilized in several devices including swirling jet nozzle, combustion fuel gas injection chamber, etc. In the scientific field, swirling flows are typically observed in tornado and cyclone. In these geophysical flows, rotating flows are often accompanied by ascending and/or descending flows (Figure~\ref{fig:4}). These helical flow structures are considered to be relevant to the genesis and sustainment of the large-scale coherent flow structures in tornados and cyclones \citep{lil1986,lev2014}. In astrophysical flow phenomena, bipolar jets ejected from the central region of an accretion disk are considered to be helical. These outflows vertical to the disk are typically accompanied by circumferential flows associated with the azimuthal rotation in the accretion disk.

\begin{figure}
\centering
\includegraphics[scale=0.8]{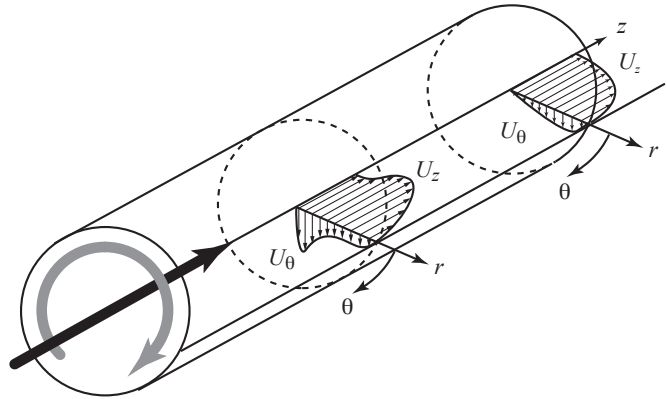}
\caption{
Configuration of swirling flow in a straight pipe.
\label{fig:5}
}
\end{figure}

	One of the prominent features of swirling jets is a dent of the jet-velocity profile near the central axis. In several types of swirling flows, it is ubiquitously observed that the axial velocity is decelerated in the center region as compared with the counterpart in the outer core region. The mean axial and circumferential velocity profiles in turbulent swirling flow in a straight pipe are schematically depicted in Figure~\ref{fig:5}. In the upstream region near the inlet, where the swirling motion is injected near the central region, the mean axial velocity shows a dent profile near the central axis. This dent profile decays towards the downstream region and finally becomes a flat profile observed in the usual turbulent pipe flow without swirl.

The deceleration of axial velocity in swirling flow can be easily understood in the laminar flow case. We consider cylindrical coordinate system $(r,\theta,z)$ with the swirling axis being in the $z$ direction. For a stationary axisymmetric swirling flow, we assume that the radial flow $u_r$ is negligibly smaller than the circumferential and axial velocities, $u_\theta$ and $u_z$. In this case, the radial momentum balance is achieved by that between the centripetal acceleration and the radial pressure gradient as
\begin{equation}
	- \frac{(u_\theta)^2}{r} 
	= - \frac{1}{\rho} \frac{\partial p}{\partial r}
	\label{eq:rad_pressure_bal}
\end{equation}
($\rho$: fluid density). Integrating (5.1) with respect to $r$ from the axis center $r=0$ to a radial position $r = r_\ast$ in the outer periphery region, and differentiate with respect to $z$, we have
\begin{equation}
	\left. {\frac{\partial p}{\partial z}} \right|_{r=0}
	- \left. {\frac{\partial p}{\partial z}} \right|_{r=r_\ast}
	= - \rho \int_{0}^{r_\ast} \frac{\partial}{\partial z} 
		\frac{(u_\theta)^2}{r} dr.
	\label{eq:int_dif_pressure_bal}
\end{equation}
This shows that the axial gradients of pressure are different between the center ($r=0$) and the outer peripheral point ($r=r_\ast$), depending on the magnitude of the circumferential velocity $u_\theta$ there. On the other hand, the axial balance of the momentum is given by
\begin{equation}
	u_z \frac{\partial u_z}{\partial z}
	= - \frac{1}{\rho} \frac{\partial p}{\partial z}.
	\label{eq:axial_mt_bal}
\end{equation}
With (\ref{eq:int_dif_pressure_bal}) this reads
\begin{eqnarray}
	\left. {\frac{\partial}{\partial z} \frac{1}{2}(u_z)^2} \right|_{r=0}
	&=& - \left. {
		\frac{1}{\rho} \frac{\partial p}{\partial z} 
	} \right|_{r=0}
	\nonumber\\
	&=& - \left. {
		\frac{1}{\rho} \frac{\partial p}{\partial z} 
		} \right|_{r=r_\ast}
	+ \int_{0}^{r_\ast} \frac{\partial}{\partial z} 
			\frac{(u_\theta)^2}{r} dr
	\nonumber\\
	&=& \left. {
		\frac{\partial}{\partial z} \frac{1}{2} (u_z)^2 
		} \right|_{r=r_\ast}
	+ \int_{0}^{r_\ast} \frac{\partial}{\partial z} 
			\frac{(u_\theta)^2}{r} dr.
	\label{eq:axial_vel_en_change}
\end{eqnarray}
This means that there is a difference between the center and periphery regions of the evolution of the axial velocity. The difference depends on the axial evolution of the circumferential velocity profile through the second term in the final r.h.s. of (\ref{eq:axial_vel_en_change}). If the swirling motion is driven in the center and inner core regions of the pipe at inlet, and decays in the core region, the second or $(u_\theta)^2$-related term in (\ref{eq:axial_vel_en_change}) is negative:
\begin{equation}
	\int_{0}^{r_\ast} \frac{\partial}{\partial z} 
		\frac{(u_\theta)^2}{r} dr < 0.
	\label{eq:axial_dif_swirl_neg}
\end{equation}
In this case, we have
\begin{equation}
	\left. {\frac{\partial}{\partial z} \frac{1}{2}(u_z)^2} \right|_{r=0}
	< \left. {
		\frac{\partial}{\partial z} \frac{1}{2} (u_z)^2 
		} \right|_{r=r_\ast},
	\label{eq:axial_decel_center}
\end{equation}
which means that in the downstream region, the axial velocity at the center  ($r=0$) is relatively small as compared to the velocity in the outer region ($r=r_\ast$), resulting in a dent profile in the central region. On the other hand, if the swirling motion is driven in the outer periphery region, in the main portion of the flow except for the near wall, the swirling velocity will increase along the axis. Then the second term in the final line of (\ref{eq:axial_vel_en_change}) is positive. In this case, the axial velocity at the center should be larger than the counterpart in the outer region, leading to the acceleration of the axial velocity in the center region. This actually occurs in the case of axially rotating pipe flow, where the rotation of the outer boundary pipe wall is the source of swirling motion of the fluid. This argument clearly shows that the decay of the swirling velocity along the axial direction is directly connected with the deceleration of the axial velocity in the center region through the pressure interaction. 

	The situation in the turbulent swirling flow is much more complicated because of the presence of the turbulent fluxes. In such a case, proper evaluation of the Reynolds stresses is of crucial importance for the accurate description and correct prediction of the swirling flow.

\subsection{Characteristics of turbulent swirling flow}
Because of its simple geometry, swirling flow in a straight pipe has been intensively investigated using experiments. These experimental studies provided elaborated data on the mean axial and circumferential velocities and the turbulence statistics in swirls \citep{kit1991,ste1995,hoe1999,gup2007}.

	The prominent features of the turbulent swirling flow in a straight pipe are as follows:

\def\theenumi{\roman{enumi}}
\def\labelenumi{(\theenumi)}
\begin{enumerate}
	
\item 
The mean axial flow is decelerated in the central axis region as compared with the axial flow in the core region (dent at the center). The degree of the deceleration increases with the swirl intensity (the scaled axial flux of the angular momentum);

\item
The swirl intensity decays exponentially with the downstream or axial distance from a reference point.

\end{enumerate}

Here, the swirl intensity $S_{\rm{W}}$ is defined by the axial flux of the angular momentum scaled by the axial flux of the linear momentum as
\begin{equation}
	S_{\rm{W}} 
	= \frac{\int_0^a 2\pi r^2 U_\theta U_z dr}{\pi a^3 U_{\rm{m}}^2},
	\label{eq:swirl_int_def}
\end{equation}
where $a$ is the radius of the pipe and $U_{\rm{m}}$ is the bulk velocity defined by $U_{\rm{m}} = \int_0^a 2\pi U_z r dr / \pi a^2$.

\begin{figure}
\centering
\includegraphics[scale=1.1]{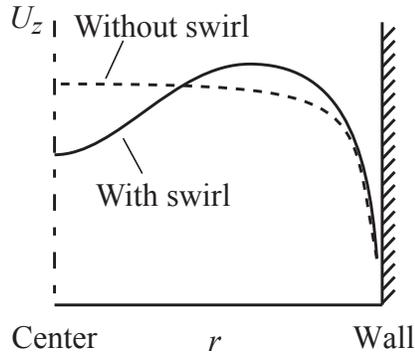}
\caption{
Mean axial velocity profile with and without swirl.
\label{fig:6}
}
\end{figure}

	The above two features are related to each other. The first feature: dent in the axial velocity at center is schematically depicted in Figure~\ref{fig:6}.

	Reversal of axial mean flow occurs in the central axis region for strong swirl intensity case. This reversal is a very important subject in the engineering application, since flow direction associated with the  combustion, fuel supply, etc.\ actually changes at some points of the flow. The condition of  axial flow reversal was investigated in variational analysis with the aid of the concept of helicity as the constraint of the variational calculus \citep{yos2001,yok2004}. It was shown theoretically that if the mean-flow helicity is larger than the critical helicity, which is proportional to the square of the mean axial momentum flux, the mean axial velocity at the central axis is reversed.

\subsection{Conventional turbulence model for swirling flow in a straight pipe\label{sec:5.3}}
	It has been well known that the standard eddy-viscosity-type $K-\varepsilon$ model would poorly reproduce both of features (i) and (ii). The standard $K-\varepsilon$ model is recovered from (\ref{eq:rey_strs_anal_result}) with the third or helicity-related term dropped. The standard $K-\varepsilon$ model applied to turbulent swirling flow in a pipe just gives a flat mean axial velocity profile, which is observed in the axial velocity profile of a turbulent pipe flow without swirling motion (Figure~\ref{fig:6}). This is because, in the presence of fully developed turbulence, the eddy-viscosity effect is so strong that no inhomogeneous mean velocity structures are allowed to exist except for the boundary layer in the vicinity of wall where the viscosity dominates nonlinearity. In order to alleviate this deficiency of the standard $K-\varepsilon$ model and to reproduce the dent axial velocity profile in the framework of the $K-\varepsilon$ model, some modifications of the model have been proposed in the engineering field. In this modified $K-\varepsilon$ model, the model constant associated with the $r$-$\theta$ component of the Reynolds stress, ${\cal{R}}_{r\theta}$, was artificially put extremely small such as $C_\mu = 0.09 \to 0.001$. Then, the dent profile of the mean axial velocity was retained in the model calculation \citep{kob1987}. However, from the requirement of the universality of model constants, which any good turbulence model should satisfy, this treatment is ad hoc and is not preferable at all.

\subsection{Analysis of turbulent swirling flow with the helicity turbulence model}
We apply the present helicity turbulence model ($K-\varepsilon-H$ model) to the turbulent swirling flow in a straight pipe. In the model, in addition to the mean velocity equation (\ref{eq:mean_vel_eq}) with the transport coefficients $\nu_{\rm{T}}$ (\ref{eq:nuT_model_Keps}) and $\mbox{\boldmath$\Gamma$}$ (\ref{eq:Gamma_model_KepsH}), the three transport equations of the turbulent statistical quantities, $K$ (\ref{eq:K_model_eq}), $\varepsilon$ (\ref{eq:eps_model_eq}), and $H$ (\ref{eq:H_model_eq}), are simultaneously solved. Then the mean fields and turbulence fields are self-consistently determined.

\begin{figure}
\centering
\includegraphics[scale=1.0]{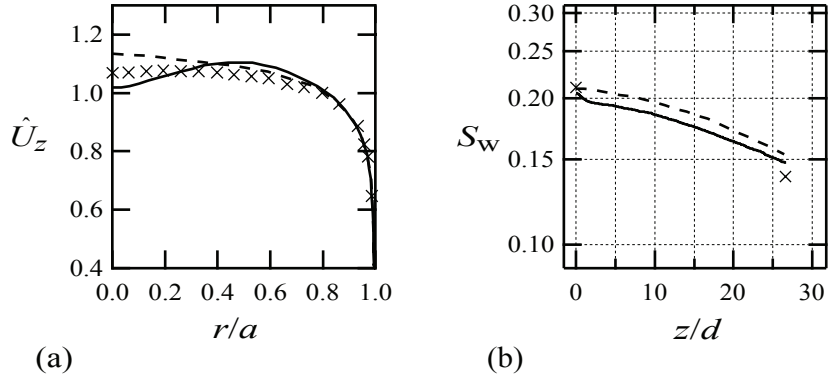}
\caption{
Radial profile of the mean axial velocity and axial decay of the swirl intensity. (a) Axial velocity normalized by the bulk velocity $\hat{U}_z = U_z/U_{\rm{M}}$, and (b) swirl intensity $S_W$. $\times$: experiment, $- - -$: standard $K-\varepsilon$ model, $--$: helicity ($K-\varepsilon-H$) model. Redrawn from \citet{yok1993}.
\label{fig:7}
}
\end{figure}

	In Figure~\ref{fig:7}, we plot the mean axial velocity $U_z$ profile and the decay of the swirl intensity $S_{\rm{W}}$ calculated by the $K-\varepsilon-H$ model in comparison with the counterparts by experiments and by the standard $K-\varepsilon$ model. As was referred to in \S~\ref{sec:5.3}, the dent profile of the mean axial velocity in the center region cannot be reproduced at all with the standard $K-\varepsilon$ model. In marked contrast, such a dent profile is naturally reproduced with the present $K-\varepsilon-H$ model without resorting to any artificial adjustment of the model constants. The exponential decay of the swirl intensity $S_{\rm{W}}$ is also more properly reproduced by the helicity ($K-\varepsilon-H$) model than the standard $K-\varepsilon$ model.

	These results clearly show the relevance of the helicity effect to treating the flow evolution of turbulent swirling flow. The inhomogeneous turbulent helicity coupled with the rotation and/or large-scale vorticity (anti-symmetric component of the mean velocity shear) contributes to balancing the enhanced transport due to the turbulent energy through the eddy viscosity which is coupled with the large-scale velocity strain (the symmetric component of the mean velocity shear).

\section{Large-scale flow generation by inhomogeneous helicity\label{sec:6}}
	In the previous section, the importance of helicity effect in the transport suppression was discussed. The turbulent helicity contributes to the dynamic balance between the transport enhancement and transport suppression. Turbulent helicity contributes to the transport suppression by counter-balancing the transport enhancement due to turbulent energy through the eddy viscosity. At the same time, turbulent helicity may contribute to inducing a global flow structure. In this section, we discuss the role of turbulent helicity in generating a large-scale flow.

\subsection{Vortex dynamo}
Replacing the vorticity $\mbox{\boldmath$\omega$}$ by the magnetic field ${\bf{b}}$ and the viscosity $\nu$ by the magnetic diffusivity $\eta$ in the vorticity equation of fluid (\ref{eq:vort_eq_rot}), we obtain the magnetic induction equation as
\begin{equation}
	\frac{\partial {\bf{b}}}{\partial t}
	= \nabla \times ({\bf{u}} \times {\bf{b}})
	+ \eta \nabla^2 {\bf{b}}.
	\label{eq:mag_ind_eq}
\end{equation}
The magnetic induction equation (\ref{eq:mag_ind_eq}) leads to the possibility of dynamo. Corresponding to the mean vorticity induction equation (\ref{eq:mean_vort_eq}), the induction equation of the mean magnetic field ${\bf{B}}$ is written as
\begin{equation}
	\frac{\partial {\bf{B}}}{\partial t}
	= \nabla \times \left( {
		{\bf{U}} \times {\bf{B}} 
		- \eta \nabla \times {\bf{B}}
	} \right)
	+ \nabla \times {\bf{E}}_{\rm{M}}
	\label{eq:mean_B_ind_eq}
\end{equation}
with the solenoidal condition for ${\bf{B}}$:
\begin{equation}
	\nabla \cdot {\bf{B}} = 0,
	\label{eq:sol_B_eq}
\end{equation}
where, ${\bf{E}}_{\rm{M}}$ is the turbulent electromotive force defined by
\begin{equation}
	{\bf{E}}_{\rm{M}} 
	= \langle {{\bf{u}}' \times {\bf{b}}'} \rangle.
	\label{eq:emf_def}
\end{equation}
The mean magnetic induction equation (\ref{eq:mean_B_ind_eq}) leads to the large-scale magnetic field generation and sustainment through the turbulent electromotive force ${\bf{E}}_{\rm{M}}$. In spite of several fundamental difference between (\ref{eq:vort_eq_rot}) and (\ref{eq:mag_ind_eq}) \citep{mof2019}, the similarity suggests a possibility of the vortex dynamo; large-scale vorticity generation and sustainment through the turbulent vortexmotive force ${\bf{V}}_{\rm{M}} = \langle {{\bf{u}}' \times \mbox{\boldmath$\omega$}'} \rangle$ in the mean vorticity equation. Actually, several studies have been undertaken on this vortex dynamo problem \citep{kra1974,moi1983,chk1988,kho1991,kit1994}. Through these studies it has been recognized that in order to have a large-scale flow generation, we need some breakage of symmetry in Reynolds stress, such as the anisotropy, compressibility, mean flow, etc. For instance, Elperin and co-workers studied an effect of mean velocity shear on homogeneous nonhelical turbulence \citep{elp2003} and on a large-scale instability for formation of vortical structure \citep{elp2007}. Spontaneous large-scale flow pattern formation in shear-flow turbulence was numerically confirmed \citep{kap2009}. These shear-flow effects in nonhelical turbulence correspond to the anisotropic state of turbulence related to the second-order nonlinear model of the Reynolds stress. The velocity shear $\nabla {\bf{U}} = \{ {\partial U_j/\partial x_i} \}$ is divided into the rate-of-strain tensor part ${\cal{S}}_{ij}$ and vorticity tensor part $\Omega_{ij}$. The nonlinear effects concerning ${\cal{S}}_{ij}$ and $\Omega_{ij}$ are expressed by the quadratic strain- and vorticity-tensor terms, such as ${\cal{S}}_{i\ell}{\cal{S}}_{\ell j}$, ${\cal{S}}_{i\ell} \Omega_{\ell j}$, $\Omega_{i\ell} \Omega_{\ell j}$. These nonlinear models lead to secondary flows in a duct flow \citep{spe1991}. Note that these nonlinear expressions of the Reynolds stress can be derived from the higher-order $O(\delta^2)$ calculation in the TSDIA formulation without resorting to the non-mirror-symmetry \citep{yos1984,yos1993}. 

	In the following, we treat the vortex dynamo problem from the viewpoint of the inhomogeneous helicity effect \citep{yok1993,yok2016b,kle2018}. 
	
	There are some mechanisms that are similar to the presenthelicity effects. One is the so-called anisotropic kinetic alpha (AKA) effect \citep{fri1987} and the other is the $\Lambda$ effect \citep{rud1980,rud1989}. The AKA effect is a large-scale instability for the global flow generation in non-mirro-symmetric turbulence. In the AKA formulation, the turbulent Reynolds number is assumed to be small, and the Reynolds stress can be expressed by using the Taylor expansion with respect to a small and uniform mean velocity. Under these assumptions, it is shown that the mean velocity evolution is subject to an anisotropic transport coefficient coupled with the mean velocity shear. As these assumptions suggest, the AKA effect operates basically only at low-Reynolds-numbers when the eddy viscosity is negligible in the momentum equation. This gives marked contrast with the present helicity effect which operates in the strongly nonlinear turbulence. On the other hand, the $\Lambda$ effect is a contribution to the Reynolds stress arising from the anisotropy of turbulence. In this formulation, the expression of the Reynolds stress is assumed to be a linear combination of the functionals of the angular velocity $\mbox{\boldmath$\omega$}_{\rm{F}}$. The $\Lambda$ effect is similar to the present helicity effect in that the both effects operate in the strong turbulence regime. This is strong contrast with the AKA effect. However, the Reynolds stress expression in $\Lambda$ effect is given as an Ansatz. In this sense, the physics that determines the transport coefficients and their validity should be examined for each case of applications.	
	As for the further descriptions of the AKA effect and the $\Lambda$ effect, the reader is referred to Appendix of \citet{yok2016b}.

	Substituting the expression of the vortexmotive force ${\bf{V}}_{\rm{M}}$ (\ref{eq:vmf_anal_result}) into the mean vorticity equation (\ref{eq:mean_vort_eq}), we have
\begin{eqnarray}
	\frac{\partial \mbox{\boldmath$\Omega$}}{\partial t}
	&=& \nabla \times \left[ {
		{\bf{U}} \times (\mbox{\boldmath$\Omega$} 
		+ 2 \mbox{\boldmath$\omega$}_{\rm{F}})
		- \nu \nabla \times \mbox{\boldmath$\Omega$}
	} \right]
	+ \nabla \times {\bf{V}}_{\rm{M}}
	\nonumber\\
	&=& \nabla \times \left[ {
		{\bf{U}} \times 
		(\mbox{\boldmath$\Omega$} 
		+ 2 \mbox{\boldmath$\omega$}_{\rm{F}})
	} \right] 
	- \nabla \times \left[ {
		(\nu + \nu_{\rm{T}}) \nabla \times \mbox{\boldmath$\Omega$}
	} \right]
	+ {\bf{I}}_{\rm{V}}.
	\label{eq:mean_vort_vmf_substituted}
\end{eqnarray}
Here, the second term represents the enhanced transport due to turbulence. The effective viscosity is enhanced from $\nu$ to $\nu + \nu_{\rm{T}}$ due to the eddy viscosity $\nu_{\rm{T}}$. The third term ${\bf{I}}_{\rm{V}}$ is defined by
\begin{equation}
	{\bf{I}}_{\rm{V}}
	= \nabla \times \left\{ {
		- D_\Gamma (\mbox{\boldmath$\Omega$}
		+ 2\mbox{\boldmath$\omega$}_{\rm{F}})
	- \left[ {
		(\mbox{\boldmath$\Omega$} 
	+ 2 \mbox{\boldmath$\omega$}_{\rm{F}}) \cdot \nabla
    } \right] \mbox{\boldmath$\Gamma$}
	} \right\},
	\label{eq:Iv_def}
\end{equation}
which denotes the possible vorticity induction due to the inhomogeneous helicity effect [$\mbox{\boldmath$\Gamma$} \propto \nabla H$ (\ref{eq:Gamma_anal_exp})]. The sign of ${\bf{I}}_{\rm{V}}$ can be positive or negative. The sign depends on the spatial distribution of the turbulent helicity, through the signs of $D_\Gamma = \nabla^2 H$ and $(\mbox{\boldmath$\Omega$}_\ast \cdot \nabla) \mbox{\boldmath$\Gamma$}$, where $\mbox{\boldmath$\Omega$}_\ast$ is the mean absolute vorticity defined by $\mbox{\boldmath$\Omega$}_\ast = \mbox{\boldmath$\Omega$} + 2 \mbox{\boldmath$\omega$}_{\rm{F}}$. In the case of negative ${\bf{I}}_{\rm{V}}$, the turbulent helicity effect works for enhancing the eddy viscosity effect.

	At the developing stage of the large-scale vorticity, where $|\mbox{\boldmath$\Omega$}| \ll |2 \mbox{\boldmath$\omega$}_{\rm{F}}|$, the induction due to the inhomogeneous helicity can be approximated as
\begin{eqnarray}
	{\bf{I}}_{\rm{V}}
	&=& \nabla \times \left[ {
		- 2 D_\Gamma \mbox{\boldmath$\omega$}_{\rm{F}}
		- \left( {
			2 \mbox{\boldmath$\omega$}_{\rm{F}} \cdot \nabla
		} \right) \mbox{\boldmath$\Gamma$}
	} \right]
	\nonumber\\
	&=& -2 (\nabla D_\Gamma) \times \mbox{\boldmath$\omega$}_{\rm{F}}
	- 2 \nabla \times \left[ {
		(\mbox{\boldmath$\omega$}_{\rm{F}} \cdot \nabla)
		\mbox{\boldmath$\Gamma$}
	} \right].
	\label{eq:Iv_omegaF_dominant}
\end{eqnarray}

\subsection{Simple argument on the origin of the flow generation due to helicity effect}
\begin{figure}
\centering
\includegraphics{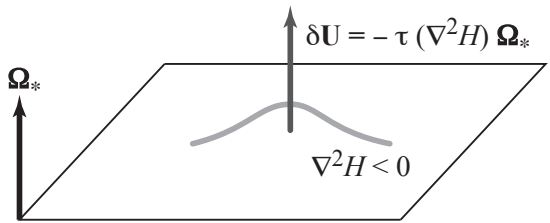}
\caption{
Possible physical origin of the global flow induction due to the inhomogeneous helicity effect.
\label{fig:8}
}
\end{figure}

Uncurling the mean vorticity equation (\ref{eq:mean_vort_vmf_substituted}), we obtain
\begin{equation}
	\frac{\partial {\bf{U}}}{\partial t}
	= {\bf{U}} \times (\mbox{\boldmath$\Omega$} 
	+ 2 \mbox{\boldmath$\omega$}_{\rm{F}})
	- (\nu + \nu_{\rm{T}}) \nabla \times \mbox{\boldmath$\Omega$}
	+ {\bf{V}}_{\rm{M}}
	+ \nabla \varphi,
	\label{eq:mean_vel_uncurled_eq}
\end{equation}
where $\varphi$ is a potential field determined by the boundary conditions whose detail is not argued here. With the turbulent vortexmotive force ${\bf{V}}_{\rm{M}}$ (\ref{eq:vmf_anal_result}), the mean velocity induced by turbulence, $\delta{\bf{U}}$, may be approximated by time integration of (\ref{eq:mean_vel_uncurled_eq}) as
\begin{equation}
	\delta {\bf{U}}
	\sim - \tau D_\Gamma \mbox{\boldmath$\Omega$}_\ast
	\sim - \tau (\nabla^2 H) \mbox{\boldmath$\Omega$}_\ast.
	\label{eq:mean_vel_laplaceH_Omega}
\end{equation}
This implies that a global flow $\delta{\bf{U}}$ is induced in the presence of inhomogeneous turbulent helicity represented by the Laplacian of the turbulent helicity $\nabla^2 H = D_\Gamma$. The flow is parallel to the mean absolute vorticity $\mbox{\boldmath$\Omega$}_\ast (= \mbox{\boldmath$\Omega$} + 2\mbox{\boldmath$\omega$}_{\rm{F}})$ for $\nabla^2 H <0$, and antiparallel for $\nabla^2 H>0$. Laplacian of a quantity represents how much the quantity is locally more or less distributed than the surroundings. It follows from (\ref{eq:mean_vel_laplaceH_Omega}) that a mean flow $\delta {\bf{U}}$ is generated in the direction parallel to the mean absolute vorticity $\mbox{\boldmath$\Omega$}_\ast$ where the turbulent helicity is locally more distributed ($\nabla^2 H <0$), and antiparallel to $\mbox{\boldmath$\Omega$}_\ast$ where the turbulent helicity is less distributed than the surroundings ($\nabla^2 H>0$). In this sense, this global flow is induced by the mean absolute vorticity (relative vorticity and rotation) with consuming the turbulent helicity redundant over that of the surroundings (Figure~\ref{fig:8}). Since the induced global velocity is parallel (or antiparallel) to the original mean absolute vorticity $\mbox{\boldmath$\Omega$}_\ast$ for $\nabla^2 H<0$ (or $\nabla^2 H>0$), a positive (or negative) mean-field helicity ${\bf{U}} \cdot \mbox{\boldmath$\Omega$}$ is generated. In this sense, back scattering of the helicity from smaller to larger scales occurs. Is this an inverse cascade of the helicity with constant fluxes? What is the values of parameter in what regime for this phenomenon? These points should be further examined in the future.

\subsection{Numerical validation of global flow generation due to helicity effect\label{sec:6.3}}
With the aid of the multiple-scale renormalized perturbation expansion method, the expression of the Reynolds stress was obtained as (\ref{eq:rey_strs_anal_result}). The essential ingredients of the helicity effect in (\ref{eq:rey_strs_anal_result}) are inhomogeneous turbulent helicity represented by $\mbox{\boldmath$\Gamma$} (\propto \nabla H)$ and the antisymmetric part of the mean velocity shear represented by $\mbox{\boldmath$\Omega$}_\ast (= \nabla \times {\bf{U}} + 2 \mbox{\boldmath$\omega$}_{\rm{F}})$. We check the validity of the Reynolds-stress expression (\ref{eq:rey_strs_anal_result}) with the aid of direct numerical simulations (DNSs) of a turbulent flow with inhomogeneous helicity (Yokoi \& Brandenburg 2016). For this purpose, we adopt a numerical setup of helical turbulence in a box in Cartesian coordinate system $(x,y,z)$ with imposed rotation $\mbox{\boldmath$\omega$}_{\rm{F}}$, whose rotation axis in the $y$ direction:
\begin{equation}
	\mbox{\boldmath$\omega$}_{\rm{F}}
	= (\omega_{{\rm{F}}x}, \omega_{{\rm{F}}y}, \omega_{{\rm{F}}z})
	= (0, \omega_{\rm{F}}, 0)
	\label{eq:omegaF_setup}
\end{equation}
as in Figure~\ref{fig:10}. The turbulent helicity $H = \langle {{\bf{u}}' \cdot \mbox{\boldmath$\omega$}'} \rangle$ is externally imposed by sinusoidal forcing.

\begin{figure}
\centering
\includegraphics[scale=0.8]{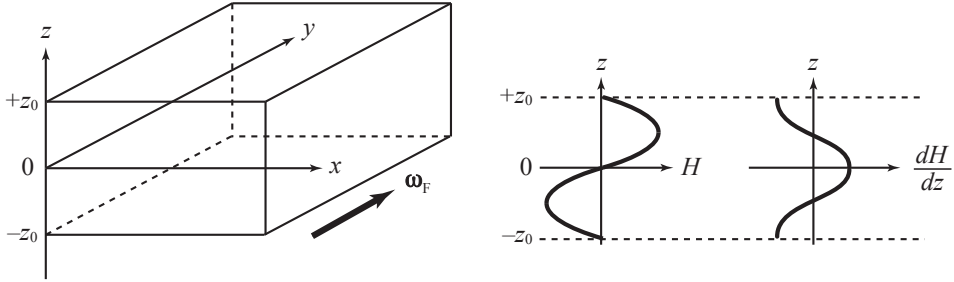}
\caption{
Setup of DNSs for the validation of helicity effect. A periodic box configuration with imposed rotation $\mbox{\boldmath$\omega$}_{\rm{F}} = (0, \omega_{\rm{F}}, 0)$ (left), and the spatial distributions of the imposed turbulent helicity $H$ and its gradient $dH/dz$ (right).
\label{fig:9}
}
\end{figure}

	In real geophysical and astrophysical flow phenomena, helicity and its inhomogeneity are supplied by the helicity production arising from mean-field inhomogeneities through the field configuration [see (\ref{eq:H_model_eq})] and/or the boundary conditions [related to the diffusion terms in (\ref{eq:K_model_eq}) and (\ref{eq:H_model_eq})]. For example, rotation coupled with inhomogeneity such as energy inhomogeneity, density stratification, asymmetry of the inertial wave propagation, etc. may be the representative mechanism that produces the turbulent helicity. Here, for the sake of simplicity of the numerical setup, we adopt the external forcing for the mechanism supplying helicity into turbulence.

	At the initial state, we do not have any mean or large-scale velocity (${\bf{U}} = 0$) at all in our setup. It follows from (\ref{eq:rey_strs_anal_result}), at the early stage of development, the $y$-$z$ component of the Reynolds stress, ${\cal{R}}_{yz}$, is given by
\begin{equation}
	{\cal{R}}_{yz} 
	= \langle {u'_y u'_z} \rangle
	= \eta 2 \omega_{{\rm{F}}} \frac{\partial H}{\partial z},
	\label{eq:Ryz_eta_omegaF_delH}
\end{equation}
where
\begin{equation}
	\eta = C_\eta \tau \ell^2
	= C_\eta \frac{K}{\varepsilon} \frac{K^3}{\varepsilon^2}
	\label{eq:eta_model}
\end{equation}
($C_\eta$: model constant).

	Once the mean flow starts being generated by the helicity effect, the Reynolds stress ${\cal{R}}_{yz}$ is expressed as
\begin{equation}
	{\cal{R}}_{yz} 
	= \langle {u'_y u'_z} \rangle
	= - \nu_{\rm{T}} \frac{\partial U_y}{\partial z}
	+ \eta 2 \omega_{{\rm{F}}} \frac{\partial H}{\partial z}.
	\label{eq:Ryz_nuT_delH_balance}
\end{equation}
Here we have assumed that the generated mean relative vorticity $\mbox{\boldmath$\Omega$} = \nabla \times {\bf{U}}$ is still much smaller than the imposed rotation ($|\mbox{\boldmath$\Omega$}| \ll 2 |\mbox{\boldmath$\omega$}_{\rm{F}}|$) to be neglected in the generation term due to the helicity. The first or $\nu_{\rm{T}}$-related term in (\ref{eq:Ryz_nuT_delH_balance}) represents the eddy-viscosity effect. The second or helicity-gradient-related term in (\ref{eq:Ryz_nuT_delH_balance}) represents the large-scale velocity generation due to inhomogeneous helicity. If the eddy viscosity $\nu_{\rm{T}}$ term and the helicity term are balanced with each other in the statistically stationary state, we have ${\cal{R}}_{yz} = \langle {u'_y u'_z} \rangle \simeq 0$. In this case, we have
\begin{equation}
	U_y \simeq \frac{\eta}{\nu_{\rm{T}}} 2 \omega_{{\rm{F}}} H.
	\label{eq:U_omega_DelH}
\end{equation}
This implies that, in such a balanced state, a mean flow ${\bf{U}}$ is induced in the direction of the imposed rotation $\mbox{\boldmath$\omega$}_{\rm{F}}$. The induced flow ${\bf{U}}$ is parallel to the rotation $\mbox{\boldmath$\omega$}_{\rm{F}}$ for the positive helicity region ($H > 0$), and antiparallel for the negative helicity region ($H < 0$). Note that inhomogeneity of the turbulent helicity is indispensable for the large-scale flow to be generated. Actually, no large-scale flow generation is observed if we impose a uniform turbulent helicity.

	In Figure~\ref{fig:10}, we present the temporal evolutions of the large-scale flow (${\bf{U}}$) and the turbulent helicity $H$. The spatial ($z$) distribution of the contours of $U_y$ and $H$ are plotted against time. Since the turbulent helicity is externally imposed by sinusoidal forcing throughout the whole period of evolution, a stationary sinusoidal pattern (positive $H$ for $z>0$ and negative $H$ for $z<0$) is observed (top). On the other hand, we have no large-scale flow (${\bf{U}} = 0$) at the initial time (bottom). Due to the inhomogeneous helicity effect, a large-scale flow starts being generated as time proceeds (early stage) and reaches to a stationary state (developed stage).

\begin{figure}
\centering
\includegraphics[scale=1.0]{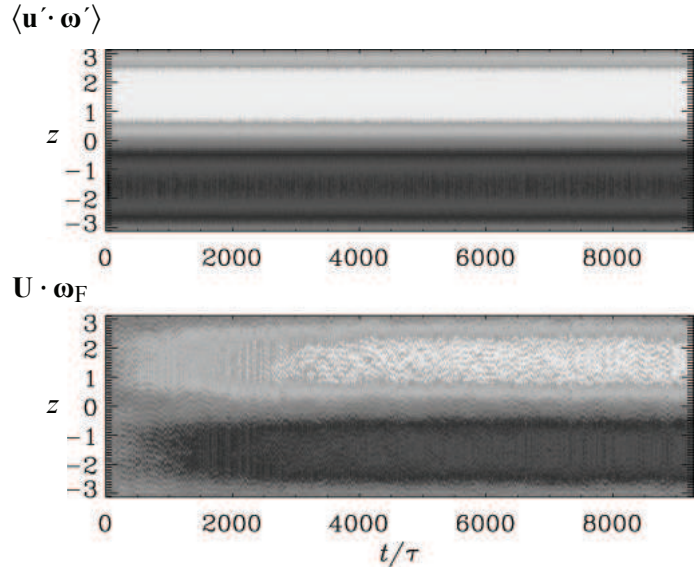}
\caption{
Temporal evolutions of turbulent helicity (top) and mean flow (bottom) with $z$ dependence. Redrawn from \citet{yok2016b}.
\label{fig:10}
}
\end{figure}

\begin{figure}
\centering
\includegraphics[scale=0.6]{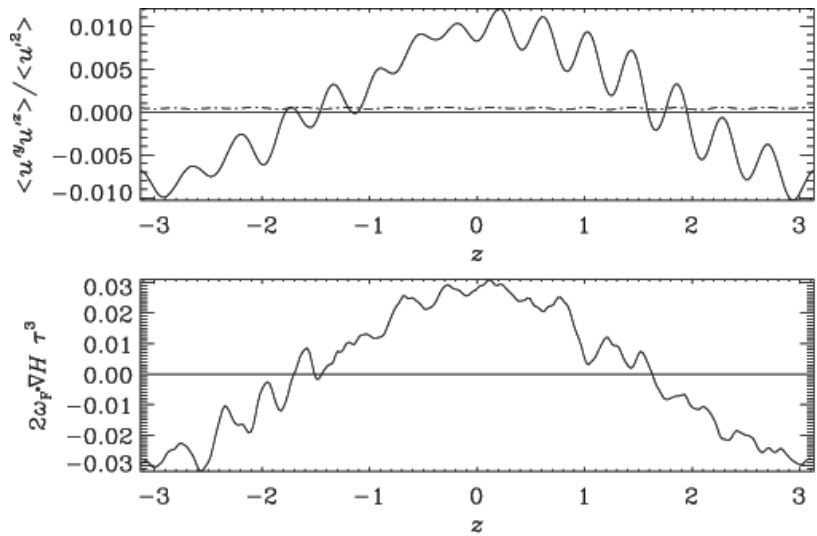}
\caption{
Reynolds stress and helicity gradient at the early stage of evolution. The $y$-$z$ component of the Reynolds stress (top) and the turbulent helicity gradient term $\omega_{\rm{F}} (\nabla H)_z$ (bottom). Averaged over time from $t/\tau = 40$ to $200$. Redrawn from \citet{yok2016b}.
\label{fig:11}
}
\end{figure}

	In Figure~\ref{fig:11}, we plot the $y$-$z$ component of the Reynolds stress, ${\cal{R}}_{yz}$ and the gradient of turbulent helicity multiplied by the rotation, $2 \omega_{{\rm{F}}} (\partial H/\partial z)$, at the early stage of the evolution (averaged over time from $t/\tau = 40$ to $200$). The spatial profile of ${\cal{R}}_{yz} = \langle {u'_y u'_z} \rangle$ is in good agreement with $2 \omega_{{\rm{F}}} (\partial H/\partial z)$. This suggests that the inhomogeneous helicity coupled with the rotation certainly works for the large-scale flow generation.

\begin{figure}
\centering
\includegraphics[scale=0.6]{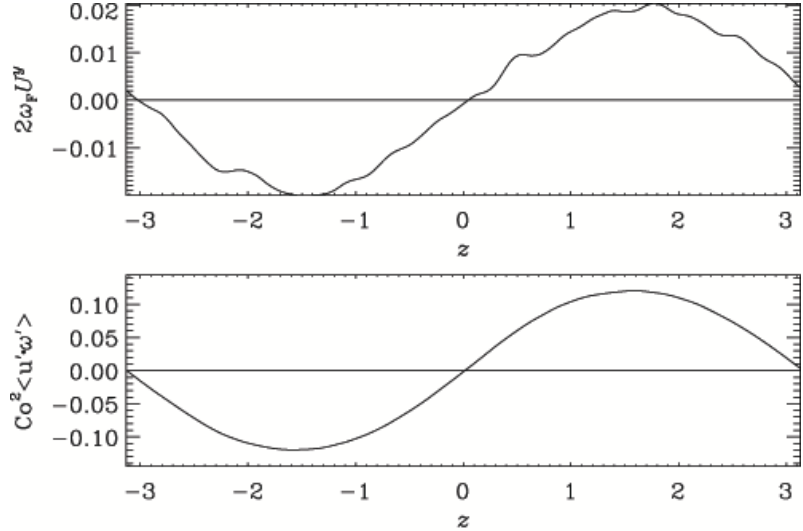}
\caption{
Induced mean velocity and turbulent helicity at the developed stage of evolution. The induced mean velocity $U_y$ (top) and the turbulent helicity $H$ (bottom). Averaged over time from $t/\tau = 0$ to $2000$. Redrawn from \citet{yok2016b}.
\label{fig:12}
}
\end{figure}

	In Figure~\ref{fig:12}, we plot the mean axial velocity $U_y$ and the turbulent helicity multiplied by the rotation, $2 \omega_{{\rm{F}}} H$, at the developed equilibrium stage (averaged over time from $t/\tau = 0$ to $2000$) against $z$. The spatial profile of the induced mean axial velocity $U_y$ is in fairly good agreement with that of the turbulent helicity $2 \omega_{{\rm{F}}} H$. This implies that, at the developed equilibrium stage, the balance between the eddy-viscosity effect $\nu_{\rm{T}} \mbox{\boldmath${\cal{S}}$} = \{ {\nu_{\rm{T}} {\cal{S}}_{ij}} \}$ and the inhomogeneous helicity effect $\mbox{\boldmath$\Gamma$} \mbox{\boldmath$\Omega$}_\ast = \{ {\Gamma_i \Omega_{\ast j}} \}$ (\ref{eq:U_omega_DelH}) is well achieved and maintained.

	We see from these results that the eddy-viscosity effect solely is not sufficient at all, and that the inhomogeneous helicity effect should be taken into account in the analysis of turbulent transport in a system with  rotating motion.

	The inhomogeneous helicity effects in flow generation and transport suppression were also confirmed by the large-eddy simulations (LESs) of rotating turbulence with externally imposed localized helicity. In the work, turbulent helicity was locally imposed by forcing with and without rotation \citep{ina2017}. A global flow generation was observed only in the case both the turbulent helicity and rotation were simultaneously imposed. Examination of each component of the Reynolds stress revealed that the counter-balancers to the eddy viscosity were the Coriolis force term and the pressure diffusion term in the Reynolds stress equations. This is natural since rotation $\mbox{\boldmath$\omega$}_{\rm{F}}$ effect can enter the turbulence dynamics only through the Coriolis-force and fluctuating-pressure terms. This result suggests that improvement of the turbulent helicity model equation (\ref{eq:H_model_eq}) may be possible by elaborated modeling of the fluctuating pressure-related terms.

\subsection{Possible applications in spherical geometry\label{sec:6.4}}

\subsubsection{Spherical surface}
In order to see the physical consequences of the induction due to the inhomogeneous helicity, we consider local Cartesian coordinate $(x,y,z)$ on the tangential spherical surface as in Figure~\ref{fig:9}. Here, $x$ denotes the azimuthal direction, $y$ the latitudinal polar direction toward the north pole, and $z$ the radial direction outward normal to the local tangential surface. We consider two limiting cases. One is the case with the length scales in the horizontal directions ($x$ and $y$) being much larger than the normal counterpart ($z$). The other is the case with the normal length scale (in the $z$ direction) being much larger than the horizontal ones (in the $x$ and $y$ directions).

\begin{figure}
\centering
\includegraphics[scale = 1.15]{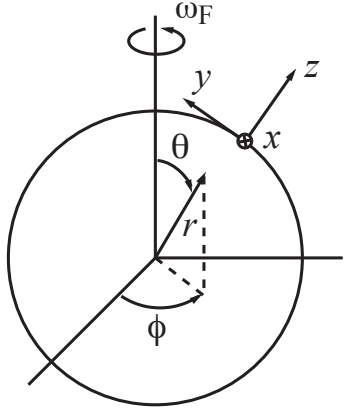}
\caption{
Local Cartesian coordinate $(x,y,z)$.
\label{fig:13}
}
\end{figure}

\paragraph{(A) Case with large horizontal scale}
	If the length scales of the horizontal directions are much larger than the normal counterpart, the gradient operator in the local Cartesian coordinate $(x,y,z)$ can be approximated as
\begin{equation}
	\nabla
	= \left( {
		\frac{\partial}{\partial x},
		\frac{\partial}{\partial y},
	\frac{\partial}{\partial z}
	} \right)
	\simeq \left( {
	0,
	0,
	\frac{\partial}{\partial z}
	} \right).
	\label{eq:local_carte_large_hor}
\end{equation}
In this case, the second term in (\ref{eq:Iv_omegaF_dominant}) vanishes. Then the vorticity induction due to the inhomogeneous helicity is given by
\begin{equation}
	{\bf{I}}_{\rm{V}}
	= - 2 (\nabla D_\Gamma) \times \mbox{\boldmath$\omega$}_{\rm{F}}.
	\label{eq:Iv_horiz}
\end{equation}
Since the angular velocity vector $\mbox{\boldmath$\omega$}_{\rm{F}}$ in the local Cartesian coordinate is given in terms of the colatitudinal polar angle $\theta$ as
\begin{equation}
	\mbox{\boldmath$\omega$}_{\rm{F}}
	= \left( {
		\omega_{{\rm{F}}x}, 
		\omega_{{\rm{F}}y},
		\omega_{{\rm{F}}z}
	} \right)
	= \left( {
		0, \omega_{\rm{F}}\sin\theta, \omega_{\rm{F}}\cos\theta
	} \right),
	\label{eq:omegaF_large_horiz}
\end{equation}
(\ref{eq:Iv_horiz}) is expressed as
\begin{equation}
	{\bf{I}}_{\rm{V}}
	= - 2 \nabla (\nabla \cdot \mbox{\boldmath$\Gamma$}) 
		\times \mbox{\boldmath$\omega$}_{\rm{F}}
	= \left( {
		2 \frac{\partial D_\gamma}{\partial z} \omega_{\rm{F}} \sin\theta,
		0,
		0
	} \right).
	\label{eq:Iv_large_horiz_omegaF}
\end{equation}
This implies that the rotation $\mbox{\boldmath$\omega$}_{\rm{F}}$ coupled with the inhomogeneous turbulent helicity may induce a large-scale vorticity in the $x$ or azimuthal direction especially in the low latitude region ($0 \ll \theta < \pi/2$). Once the large-scale vorticity is generated in the azimuthal direction, $\Omega_x$, the first term in (\ref{eq:Iv_def}) starts working and gives
\begin{equation}
	I_{{\rm{V}}y}
	\simeq \left\{ {
		\nabla \times \left[ {
		- D_\Gamma (\mbox{\boldmath$\Omega$} 
		+ 2 \mbox{\boldmath$\omega$}_{\rm{F}})
		} \right]
	} \right\}_y
	= - \frac{\partial}{\partial z} (D_\Gamma \Omega_x),
	\label{eq:Ivy_large_horiz_omegaF}
\end{equation}
inducing the latitudinal component of the large-scale vorticity, $\Omega_y$. This mechanism causes a helical large-scale vortical structure in the low latitude region at the surface of a rotating stellar and/or planetary object.

\paragraph{(B) Case with large normal scale}
	If the length scale in the direction normal to the tangential surface is much larger than the counterparts in the horizontal directions, the gradient operator is approximated as
\begin{equation}
	\nabla
	= \left( {
		\frac{\partial}{\partial x},
		\frac{\partial}{\partial y},
		\frac{\partial}{\partial z}
	} \right)
	\simeq \left( {
		\frac{\partial}{\partial x},
		\frac{\partial}{\partial y},
		0
	} \right).
	\label{eq:local_carte_large_norm}
\end{equation}
In this case, the vector $\mbox{\boldmath$\Gamma$}$ related to the helicity gradient in (\ref{eq:Iv_def}) is written as
\begin{equation}
	\mbox{\boldmath$\Gamma$}
	= \left( {
		\Gamma_x, \Gamma_y, 0
	} \right)
	\simeq \left( {
		\frac{\partial H}{\partial x},
		\frac{\partial H}{\partial y},
		0
	} \right).
	\label{eq:Gamma_large_norm}
\end{equation}
At the early stage of the vorticity generation ($|\mbox{\boldmath$\Omega$}| \ll |2\mbox{\boldmath$\omega$}_{\rm{F}}|$), we see from (\ref{eq:Iv_omegaF_dominant}) that the helicity contribution to the vorticity induction is given by
\begin{eqnarray}
	{\bf{I}}_{\rm{V}}
	&=& \nabla \times \left[ {
		- 2 D_\Gamma \mbox{\boldmath$\omega$}_{\rm{F}}
		- \left( {
			2 \mbox{\boldmath$\omega$}_{\rm{F}} \cdot \nabla
		} \right) \mbox{\boldmath$\Gamma$}
	} \right]
	\nonumber\\
	&=& \left( {
		-2 \frac{\partial D_\Gamma}{\partial y} \omega_{\rm{F}} \cos\theta,
		2 \frac{\partial D_\Gamma}{\partial x} \omega_{\rm{F}} \cos\theta,
		-2 \frac{\partial D_\Gamma}{\partial x} \omega_{\rm{F}} \sin\theta
	} \right)
	\nonumber\\
	&&\hspace{10pt}+ \left( {
		0,
		0,
		2 \omega_{\rm{F}} \sin\theta \left( {
			\frac{\partial^2 \Gamma_x}{\partial y^2} 
			- \frac{\partial^2 \Gamma_y}{\partial x \partial y}
		} \right)
	} \right).
	\label{eq:Iv_large_norm_omegaF}
\end{eqnarray}
For the lower latitude region, this is reduced to
\begin{equation}
	{\bf{I}}_{\rm{V}}
	= \left( {
		0,
		0,
		2 \omega_{\rm{F}} \sin\theta \left( {
			- \frac{\partial D_\Gamma}{\partial y}
			+ \frac{\partial^2 \Gamma_x}{\partial y^2} 
			- \frac{\partial^2 \Gamma_y}{\partial x \partial y}
		} \right)
	} \right).
	\label{eq:Iv_norm_lower_lat}
\end{equation}
This implies that the rotation vector coupled with the inhomogeneous turbulence helicity induces a large-scale vorticity in the direction normal to the tangential surface in the lower latitude region. This may contribute to the enhancement of the genesis of the cyclone in the near equatorial region \citep{yok1993}.

\subsubsection{Angular momentum transport}
Equation~(\ref{eq:mean_vel_laplaceH_Omega}) implies that inhomogeneous spatial distribution of turbulent helicity coupled with the mean vortical motion and/or rotation can induce a large-scale flow in the direction of the mean vorticity/rotation. This can be paraphrased that the angular momentum is transported to the domain of flow acceleration and/or generation due to the inhomogeneous turbulent helicity effect. Two interesting cases are suggested below.

\paragraph{(A) Stellar angular momentum transport}
	Stellar and planetary interior convective motions play key roles in forming and sustaining form and sustain the stellar and planetary magnetic field by dynamo actions. Recent developments in helioseismology unraveled flow configurations inside the Sun: the latitudinal and radial distributions of the solar angular velocity as well as the basic patterns of the meridional circulation. It has been established that the solar azimuthal rotation represented by the solar angular velocity is faster in the near equator and near surface regions \citep{mie2005}. 
	
\begin{figure}
\centering
\includegraphics[scale=0.6]{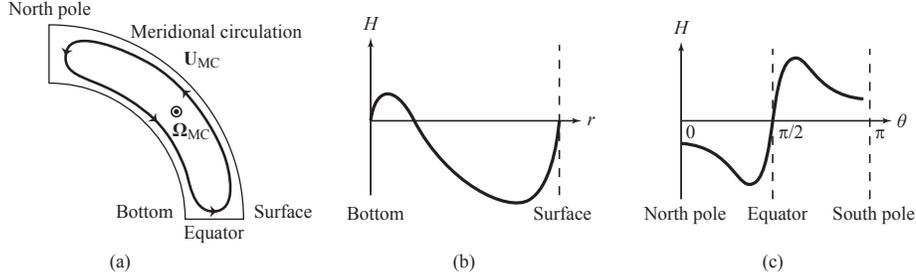}
\caption{
Angular momentum transport in the Sun viewed from the inhomogeneous helicity effect. (a) Basic pattern of the meridional circulation [${\bf{U}}_{\rm{MC}}$: meridional circulation velocity, $\mbox{\boldmath$\Omega$}_{\rm{MC}} (= \nabla \times {\bf{U}}_{\rm{MC}}$): Vorticity associated with the meridional circulation]. (b) The radial variation of the turbulent helicity. (c) The colatitudinal variation of the turbulent helicity. All figures are schematically depicted.
\label{fig:14}
}
\end{figure}

	Although the detailed pattern of the meridional circulation is still an open problem, the basic pattern of it may be considered as the one depicted in Figure~\ref{fig:14}(a). In the surface region, the meridional circulation ${\bf{U}}_{\rm{MC}}$ flows poleward direction while at some depth it returns to equatorward. Then, the mean vorticity associated with the meridional circulation $\mbox{\boldmath$\Omega$}_{\rm{MC}} (= \nabla \times {\bf{U}}_{\rm{MC}})$, is in the westward azimuthal direction.

	On the other hand, the spatial distribution of the turbulent kinetic helicity has been obtained using DNSs of the spherical shell domain mimicking the stellar and planetary convective zones \citep{dua2016}. Following the numerical results, the radial distribution of the turbulent helicity is schematically depicted in Figure~\ref{fig:14}(b). At the bottom boundary of the convective zone, the helicity is null, and near the bottom region it is positive. The positive helicity decreases with radius, and becomes negative in the shallower domain of the convective zone. The magnitude of negative helicity decreases as approaching the surface, and finally becomes null at the surface boundary. This spatial distribution of helicity is naturally understood with the helicity generation mechanisms due to the inhomogeneity along the rotation $(\mbox{\boldmath$\omega$}_{\rm{F}} \cdot \nabla) K$ (\ref{eq:inhomo_K_along_rot}) discussed in \S~\ref{sec:hel_turb_model}. If a lump of fluid with angular momentum approaches the outer boundary, it is expanded in the horizontal direction due to impinging. Because of the local angular momentum conservation, the rotation of the lump decreases, leading to a negative swirl motion as compared to that of the ambients. This is alternatively expressed by the inhomogeneity along the rotation term.

	We apply the inhomogeneous helicity effect on the mean velocity [(\ref{eq:rey_strs_anal_result}) with (\ref{eq:Gamma_simpl_exp})] to the problem of angular momentum transport inside the Sun. Let us consider the angular-momentum transport in spherical polar coordinate system $(r,\theta,\phi)$. The angular momentum around the rotation axis, $L$, is defined by
\begin{equation}
	L = \Gamma r^2 \omega_{\rm{F}} + \Lambda r U_\phi
	\label{eq:ang_mt_def}
\end{equation}
with $\Lambda = \sin\theta$. The transport of $L$ is subject to 
\begin{equation}
	\frac{\partial}{\partial t} \rho L
	+ \nabla \cdot \left( {\rho {\bf{F}}_L} \right)
	= 0.
	\label{eq:transport_L}
\end{equation}
Here $\rho$ is the density and ${\bf{F}}_L$ is the vector flux of the angular momentum, expressed as 
\begin{equation}
	F_{Lr} = L U_r + r \Lambda {\cal{R}}_{r\phi},
	\label{eq:FLr_def}
\end{equation}
\begin{equation}
	F_{L\theta} = L U_\theta + r \Lambda {\cal{R}}_{\theta\phi},
	\label{eq:FLtheta_def}
\end{equation}
where $\mbox{\boldmath${\cal{R}}$} (= \{ {{\cal{R}}_{ij}} \} = \{ {\langle {u'_i u'_j} \rangle} \})$ is the Reynolds stress.

	In the axisymmetric case ($\partial/\partial \phi$), the inhomogeneous helicity contribution to the Reynolds stress can be written as
\begin{equation}
	{\cal{R}}_{\theta\phi}^{(H)}
	= + \frac{1}{r} \frac{\partial H}{\partial \theta} \left( {
		\frac{1}{r} \frac{\partial}{\partial r} r U_\theta
		- \frac{1}{r} \frac{\partial U_r}{\partial \theta}
	} \right),
	\label{R_thetaphi_H}
\end{equation}
\begin{equation}
	{\cal{R}}_{r\phi}^{({H})}
	= + \frac{\partial H}{\partial r} \left( {
		\frac{1}{r} \frac{\partial}{\partial r} r U_\theta
	- \frac{1}{r} \frac{\partial U_r}{\partial \theta}
	} \right).
	\label{eq:R_rphi_H}
\end{equation}
In the presence of meridional circulation in the poleward direction at the solar surface, the toroidal mean vorticity is negative in the northern hemisphere as
\begin{equation}
	\frac{1}{r} \frac{\partial}{\partial r} r U_\theta 
	\left\{ 
	\begin{array}{ll}
  	< 0 & (\mbox{Northern hemisphere})\\
	> 0 & (\mbox{Southern hemisphere})
	\end{array}
	\right.
	\label{eq:solar_Omega_phi}
\end{equation}
On the other hand, the turbulent helicity in the northern hemisphere is expected to be more negative in the shallow region (except for the vicinity of the surface) than the deeper region as
\begin{equation}
	\frac{\partial H}{\partial r} 
	\left\{ {
	\begin{array}{ll}
	< 0 & (\mbox{Northern hemisphere})\\
	> 0 & (\mbox{Southern hemisphere})
	\end{array}
	} \right.
	\label{eq:solar_dHdr}
\end{equation}
At the same time, the colatitudinal gradient of turbulent helicity is negative in both hemispheres except in the vicinity of the equator where the sign of helicity changes from negative to positive [Figure~\ref{fig:14}(c)]. Then, we expect
\begin{equation}
	\frac{1}{r} \frac{\partial H}{\partial \theta} 
	\left\{ {
	\begin{array}{ll}
	< 0 & (\mbox{Northern hemisphere})\\
	< 0 & (\mbox{Southern hemisphere})
	\end{array}
	} \right.
	\label{eq:solar_dHdtheta}
\end{equation}
It follows from (\ref{eq:solar_Omega_phi})-(\ref{eq:solar_dHdtheta}) that the helicity contributions to the angular momentum fluxes are written as
\begin{equation}
	r \Lambda {\cal{R}}_{\theta\phi}^{(H)}
	\simeq + \Lambda \frac{\partial H}{\partial \theta}   \frac{1}{r} 	
		\frac{\partial}{\partial r} r U_\theta > 0,
	\label{eq:ang_mt_flux_thetaphi}
\end{equation}
\begin{equation}
	r \Lambda {\cal{R}}_{r\phi}^{(H)}
	\simeq + r \Lambda \frac{\partial H}{\partial r}   \frac{1}{r} 
	\frac{\partial}{\partial r} r U_\theta
	> 0.
	\label{eq:ang_mt_flux_rphi}
\end{equation}
Equations (\ref{eq:ang_mt_flux_thetaphi}) and (\ref{eq:ang_mt_flux_rphi}) show that the helicity contributions to the $\theta$-$\phi$ and $r$-$\phi$ components are positive. This means that the helicity effect works for accelerating the azimuthal velocity by transporting the angular momentum towards shallower and lower latitude (near equator) region. This is in agreement with the results of helioseismic observations. Preliminary studies utilizing direct numerical simulations (DNSs) of a spherical shell mimicking the solar convective zone shows that the magnitudes of the helicity terms, $|(\nabla H) \mbox{\boldmath$\Omega$}| = \{ {|{(\partial H/\partial x_i) \Omega_j}|} \}$, with the parameters for the solar interior, are comparable to those of ${\cal{R}}_{r\phi}$ and ${\cal{R}}_{\theta\phi}$ \citep{mie2017}. This suggests that this inhomogeneous helicity effect should be further investigated in the solar interior context.

	More intuitively, we can argue the role of inhomogeneous helicity effect on the angular momentum transport in the Sun as follows. We see from (\ref{eq:mean_vel_laplaceH_Omega}) that the mean velocity variation due to the helicity effect is given as
\begin{equation}
	\delta {\bf{U}}_{\rm{MC}}
	\sim - \tau (\nabla^2 H) \mbox{\boldmath$\Omega$}_{\rm{MC}}.
	\label{eq:delU_MC_Omega_MC}
\end{equation}
Since $\nabla^2 H$ is positive in the shallow region [Figure~10(a)], the variation of the mean velocity $\delta {\bf{U}}_{\rm{MC}}$ (\ref{eq:delU_MC_Omega_MC}) is expected to be antiparallel to the mean vorticity due to the meridional circulation, $\mbox{\boldmath$\Omega$}_{\rm{MC}}$. We see that this inhomogeneous helicity effect works for the acceleration of the azimuthal angular velocity in the shallow convective zone, which again matches the helioseismic observation.

\begin{figure}
\centering
\includegraphics[scale=0.6]{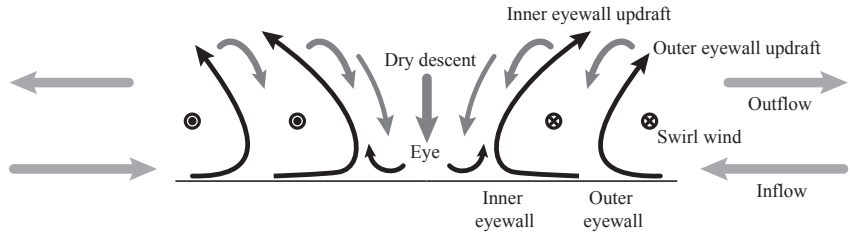}
\caption{
Flow configuration of tropical cyclone.
\label{fig:15}
}
\end{figure}

\paragraph{(B) Wind acceleration near cyclone eyewall}
	Another possibly interesting application of the inhomogeneous helicity effect is the wind acceleration in the eyewall formation region of tropical cyclone. It is observationally known that the double-eyewall structure is formed in the core of a cyclone (Figure~\ref{fig:15}), and a distinct maximum of the wind speed (tangential velocity of cyclone) is located at the radial position of each eyewall. Such double-eyewall formation and maximum wind speed at the eyewalls have been previously discussed in terms of the radial propagation of the vortex Rossby wave and its stagnation at the eyewall formation regions \citep{mon1997,hua2012,hou2010}.

	Since eyewall regions are related to inhomogeneous helicity generation, it may be possible to explain such a wind acceleration by the inhomogeneous helicity effect. The common configurations of cyclones are upward drafts at inner and outer eyewalls and dry forced descents at eye (Figure~\ref{fig:15}). The updraft and its circulated down flow constitute the azimuthal component of the mean vorticity. At the same time, presence of vertical flows in the cyclone eye region coupled with the vertical mean vorticity associated with the cyclone wind velocity indicates that a mean-flow helicity ${\bf{U}} \cdot \mbox{\boldmath$\Omega$}$ is present there. This suggests that the turbulent helicity must be highly inhomogeneous near eyewall. Such a strong inhomogeneous mean helicity naturally results in the inhomogeneous spatial distribution of the turbulent helicity $H = \langle {{\bf{u}}' \cdot \mbox{\boldmath$\omega$}'} \rangle$. Combination of the mean azimuthal vorticity and the inhomogeneous helicity near the eyewall may induce a large-scale flow in the azimuthal direction following (\ref{eq:mean_vel_laplaceH_Omega}).  
	
\section{Summary and conclusion\label{sec:7}}
Helicity contributes to the alteration of flow dynamics and fundamental statistical properties in turbulence. In contrast to energy, which represents the intensity properties of fluctuations, helicity represents structural properties of fluctuations. Incorporation of the helicity effects into turbulence statistical theory is achieved by inclusion of the cross-flow or cross-component correlation as well as the longitudinal and transverse velocity correlations (\ref{eq:gene_R_GFH}). In the configuration space, the longitudinal and transverse velocity correlations are linked to the turbulent energy, while the cross correlation is linked to the turbulent helicity. Since helicity is a measure of breakage of mirror-symmetry, its effects are incorporated into theoretical analysis by taking into account the non-mirror-symmetric part in the expression of the fluctuation correlations in the wave-number space [(\ref{eq:gen_Rij_comp}) and (\ref{eq:gen_Rij_sol})].

	Turbulent transport is expressed by turbulent fluxes such as the Reynolds stress in the mean momentum equation and the vortexmotive force in the mean vorticity equation. In mirror-symmetric case, turbulent transport is determined solely by the intensity of fluctuation. The more intense fluctuation is, the more effective transport we have. From the viewpoint of turbulence model, the eddy viscosity represents such an enhanced transport due to turbulence. The turbulent viscosity is expressed in terms of the turbulent energy and its timescale. In non-mirror-symmetric case, helicity as well as energy enter the expressions of turbulent fluxes as the descriptor of turbulent transport. 

	With the aid of the multiple-scale renormalized perturbation expansion theory, we obtained the analytical expressions for the Reynolds stress and turbulent electromotive force. In these theoretical results, the transport coefficients are expressed in terms of the spectral correlation functions and response function (Green's function). The eddy viscosity, the transport coefficient coupled with the symmetric part of the mean velocity shear (mean velocity strain), is expressed by the spectral and time integral of the energy spectral function and the Green's functions (\ref{eq:nuT_anal_exp}). In a system with lacking reflectional symmetry, in addition to the eddy viscosity, another transport coefficient coupled with the antisymmetric part of the mean velocity shear (mean vorticity and rotation) shows up. The latter transport coefficient is expressed by the inhomogeneity of the helicity spectral function and the Green's function (\ref{eq:Gamma_anal_exp}). 

	On the basis of these analytical results, we construct a turbulence model with the structural effect incorporated through the turbulent helicity. In order to construct a self-consistent system of model equations, the mean and turbulence fields should be simultaneously solved. For this purpose, the transport equations of the turbulent energy $K$, its dissipation rate $\varepsilon$, and the turbulent helicity $H$ were  proposed ($K-\varepsilon-H$ model). This model was applied to swirling flow in a straight pipe. This helicity turbulence model could successfully reproduce the main characteristics of the swirling flow: the deceleration of the mean axial velocity and the exponential decay of the swirl intensity, which could not be reproduced at all with the standard $K-\varepsilon$ model with the eddy-viscosity representation. This numerical result shows that the eddy viscosity which enhances the effective transport too much for the swirling flow, can be successfully counter-balanced by the inhomogeneous helicity effect.

	The helicity effect was also validated by using direct numerical simulations (DNSs) of a rotating triple-periodic symmetrical box with inhomogeneous helicity externally imposed by forcing. Starting with no initial mean flow configuration, it was shown that inhomogeneous helicity coupled with rotation contributes to the induction of a global flow in the direction of the rotation vector. It was shown that the Reynolds stress shows a high correlation with the inhomogeneous helicity at the early stage of flow evolution, where the eddy-viscosity effect is weak because of the absence of mean velocity strain. Also the high correlation between the induced mean flow and the local turbulent helicity at the developed stage confirms the balancing between the eddy viscosity and inhomogeneous helicity effects. These numerical validations show that, in the non-mirror-symmetric case, the helicity effect coupled with the mean absolute vorticity (mean vorticity and rotation) should be considered in addition to the usual eddy-viscosity effect coupled with the mean velocity strain.

	From these theoretical and numerical results, we can summarize the roles of helicity in fluid turbulent transport. In non-mirror-symmetric system, the Reynolds stress is schematically expressed as
\begin{eqnarray}
	\langle {{\bf{u}}' {\bf{u}}'} \rangle
	:= &\overbrace{- \nu_{\rm{T}} \mbox{\boldmath${\cal{S}}$}}  
		^{\mbox{Eddy viscosity}}
  	&\overbrace{+ \mbox{\boldmath$\Gamma$} \mbox{\boldmath$\Omega$}_\ast}
	^{\mbox{Helicity effect}}.\\
	\label{eq:rey_strs_enhance_suppress}
	\mbox{Transport:} & \mbox{Enhancement} & \mbox{Suppression}
	\nonumber\\
	\mbox{Flow/Structure:} & \mbox{Destruction} & \mbox{Generation}
	\nonumber
\end{eqnarray}
Here, for the sake of brevity, the notations for the deviatoric and transposed component of a tensor are suppressed, $:=$ denotes ``schematically represents'', $\mbox{\boldmath${\cal{S}}$}$ is the mean velocity strain tensor, $\mbox{\boldmath$\Gamma$} \propto \nabla H$ is the gradient of helicity, and $\mbox{\boldmath$\Omega$}_\ast (= \mbox{\boldmath$\Omega$} + 2\mbox{\boldmath$\omega$}_{\rm{F}})$ is the mean absolute vorticity. The eddy viscosity $\nu_{\rm{T}}$ is determined by the turbulent energy and timescale of turbulence, while the helicity-related transport coefficient $\mbox{\mbox{\boldmath$\Gamma$}}$ is determined by the gradient of helicity, timescale and length scale of turbulence. It is worth noting that turbulence timescale and length scale are altered by the presence of strong helicity, rotation, velocity shear, non-equilibrium effect etc.\ \citep{yok2022}. Actually, the implication of the strain- and vorticity-tensor effect on the timescale and length scale of turbulence has been explored in the conventional nonlinear turbulence modeling studies \citep{spe1991}.

	In treating turbulent transport in non-mirror-symmetric turbulence, the helicity effects, which represent transport suppression and structure generation, have to be incorporated into the turbulent fluxes and their model expressions as well as the turbulent energy effect. In the present helicity effects, the direction of change in the turbulent transport and global flow generation/destruction depends on the spatial distribution of the turbulent helicity. In this sense, spatiotemporal evolution of the turbulent helicity, its production, dissipation, and transport mechanisms play a crucial role. In the real-world turbulence, such mechanisms are determined by nonlinear dynamics between the mean and turbulence fields. A self-consistent approach based on turbulence modeling described in \S~\ref{sec:turb_model} is of fundamental importance in exploring global flow generation mechanisms in nature.

	Much more attention should be paid for helicity. In particular, detailed laboratory experiments and in-situ and/or remote observations, numerical calculations of the spatiotemporal distributions of helicity (kinetic helicity, magnetic helicity, current helicity, residual helicity, cross helicity, \dots) are most required for developing our insights in turbulence and nonlinear flow phenomena.

\vspace{20pt}
\noindent{\bf Acknowlegments}

My thanks are due to the anonymous reviewers for their thoughtful comments based on deep knowledge and insight, which contributed much to the improvement of the manuscript. I would like to cordially dedicate this work to the memory of my parents: Hisashi Yokoi (1928.3.14-1999.7.18) and Taduko Yokoi (1929.10.13-2020.4.22), who had provided agapeic encouragements and led me into a scholarly life.

\backmatter

\bibliographystyle{plainnat}
\bibliography{Wiley}%





\backmatter

\printindex

\end{document}